\documentclass[english]{scrartcl}
\usepackage[T1]{fontenc}
\usepackage[latin9]{inputenc}
\usepackage{geometry}
\usepackage{float}
\usepackage{units}
\usepackage{amsmath}
\usepackage{amssymb}
\usepackage{graphicx}
\usepackage{setspace}
\usepackage{esint}
\usepackage[authoryear]{natbib}
\makeatletter

\providecommand{\tabularnewline}{\\}

\usepackage{lineno}
\usepackage{url}
\usepackage{natbib}
\bibpunct{[}{]}{;}{a}{,}{,}
\usepackage[font=small,format=plain,labelfont=bf, margin=0.5cm]{caption}
\usepackage{hyperref}

\hypersetup{colorlinks=flase,citecolor=black,linkcolor=black}
\date{}

\makeatother

\usepackage{babel}
\begin{document}

\title{Variational Downscaling, Fusion and Assimilation of Hydrometeorological States
via Regularized Estimation}

\author{A.M. Ebtehaj\textsuperscript{1,2}, E. Foufoula-Georgiou\textsuperscript{1},
\\
{\small \textsuperscript{1}Department of Civil Engineering, Saint
Anthony Falls Laboratory, University of Minnesota}\\
{\small \textsuperscript{2}School of Mathematics, University of Minnesota}}
\maketitle
\begin{abstract}
Improved estimation of hydrometeorological states from down-sampled
observations and background model forecasts in a noisy environment,
has been a subject of growing research in the past decades. Here,
we introduce a unified framework that ties together the problems of
downscaling, data fusion and data assimilation as ill-posed inverse
problems. This framework seeks solutions beyond the classic least
squares estimation paradigms by imposing proper regularizations, which
are constraints consistent with the degree of smoothness and probabilistic
structure of the underlying state. We review relevant regularization
methods in derivative space and extend classic formulations of the
aforementioned problems with particular emphasis on hydrologic and
atmospheric applications. Informed by the statistical characteristics
of the state variable of interest, the central results of the paper
suggest that proper regularization can lead to a more accurate and
stable recovery of the true state and hence more skillful forecasts.
In particular, using the Tikhonov and Huber regularizations in the
derivative space, the promise of the proposed framework is demonstrated
in static downscaling and fusion of synthetic multi-sensor precipitation
data, while a data assimilation numerical experiment is presented
using the heat equation in a variational setting.
\end{abstract}

\section{Introduction}

In parallel to the growing technologies for earth remote sensing,
we have witnessed an increasing interest to improve the accuracy of
observations and integrate them with physical models for more accurate
estimates and forecasts of environmental states. Remote sensing observations
are typically noisy and coarse-scale representations of a true state
variable of interest, lacking sufficient accuracy and resolution for
fine-scale environmental modeling. In addition, environmental predictions
are not perfect as models often suffer either from inadequate characterization
of the underlying physics or their initial conditions. Given these
limitations, several classes of estimation problems present themselves
as continuous challenges for the atmospheric, hydrologic and oceanic
science communities. These include: (1) Downscaling which refers to
the class of problems to enhance the resolution of a measured field
or produce a fine-scale representation of a coarse-scale modeled field;
(2) Data fusion, to produce an improved estimate from a suite of noisy
observations at different scales; and (3) Data assimilation which
deals with estimating initial conditions in a predictive model consistent
with the available noisy observations and the model dynamics. In this
paper, we revisit the problems of downscaling (DS), data fusion (DF),
and data assimilation (DA) focusing on a common thread between them
as inverse estimation problems. Proper regularizations and solution
methods are proposed to efficiently handle large-scale data sets while
preserving key statistical and geometrical properties of the underlying
fields of interest. Here, we only examine hydrometeorological problems
with particular emphasis on land-surface applications.

In land surface hydrologic studies, downscaling of precipitation and
soil moisture signals has received considerable attention, using a
relatively wide range of methodologies. DS methods in hydrometeorology
and climate studies generally fall into three main categories namely,
dynamic downscaling, statistical downscaling, and variational downscaling.
Dynamic downscaling often uses a regional physical model to reproduce
fine-scale details of the state of interest consistent with the large-scale
scale observations or outputs of a global circulation model \citep[e.g.,][]{[ReiEM01],[CasCPL05],[ZupDZZMH10]}.
Statistical downscaling methods encompass a large group of methods
that typically use empirical multiscale statistical relationships,
parameterized by observations or other environmental predictors, to
reproduce realizations of fine-scale fields. Precipitation and soil
moisture statistical downscaling has been mainly approached via spectral
and (multi)fractal interpolation methods, capitalizing on the presence
of a power law spectrum and a statistical self-similarity/self-affinity
in precipitation and soil moisture fields \citep[among others,][]{[lovM85],[LovS90],[GupW93],[KumF93a],[KumF93b],[PerF96],[Ven96],[WilWC98],[WilHH98],[Dei00],[kimB02],[RebFHP05],[BadDP06],[MerCKG06]}.
In variational approaches, a direct cost function is defined whose
optimal point is the desired fine-scale field which can be obtained
via using an optimization routine. Recently along this direction,
\citet{[EbtF12a]} cast the rainfall DS problem as an inverse problem
using sparse regularization to address the intrinsic rainfall singularities
and non-Gaussian statistics. This variational approach belongs to
the class of methodologies presented and extended in this paper.

The DF problem has also been a subject of continuous interest in the
precipitation science community mainly due to the availability of
rainfall measurements from multiple spaceborne (e.g., TRMM and GOES
satellites) and ground-based sensors (e.g., the NEXRAD network). The
accuracy and space-time coverage of remotely sensed rainfall are typically
conjugate variables. In other words, more accurate observations are
often available with lower space-time coverage and vice verse. For
instance, low-orbit microwave sensors provide more accurate with less
space-time coverage compared to the high-orbit geo-stationary infrared
(GOES-IR) sensors. Moreover, there are often, multiple instruments
on a single satellite platform (e.g., precipitation radar and microwave
imager on TRMM), each of which measures rainfall with different footprints
and resolutions. A wide range of methodologies including weighted
averaging, regression, filtering, and neural networks has been applied
to combine microwave and Geo-IR rainfall signals \citep[e.g.,][]{[Adletal03],[Hufetal.95],[SorH00],[Hufetal01],[HonHSG04],[Hufetal07]}.
Furthermore, a few studies have addressed methodologies to optimally
combine the products of the TRMM precipitation radar (PR) with the
TRMM microwave imager (TMI) using Bayesian inversion and weighted
least squares approaches \citep[e.g.,][]{[MasK05],[KumRCRB10]}. From
another direction, Gaussian filtering methods on Markovian tree-like
structures, the so-called scale-recursive-estimation (SRE), have been
proposed to merge spaceborne and ground-based rainfall observations
at multiple scales \citep[e.g.,][]{[GorME01],[TusaFH03],[Boc07],[VanR09],[WanLN11]},
see also \citep{[Kum99]} for soil moisture application. Recently,
using the Gaussian scale mixture probability model and an adaptive
filtering approach \citet{[EbtF11b]} proposed a fusion methodology
in the wavelet domain to merge TRMM-PR and ground-based NEXRAD measurements,
aiming to preserve the non-Gaussian structure and local extremes of
precipitation fields.

Data assimilation has played an important role in improving the skill
of environmental forecasts and has become by now a necessary step
in operational prediction models \citep[see][]{[Dal93]}. Data assimilation
amounts to integrating the underlying knowledge from the \emph{observations}
into the first guess or the \emph{background} state, typically provided
by a physical model. The goal is then to obtain the best estimate
of the current state of the system, the so called \emph{analysis}.
The analysis with reduced error is then used to forecast the state
at the next time step and so on \citep[see][for a comprehensive review]{[Dal93],[Kal03]}.
Note that, in practice, the present background state, used in each
analysis cycle, is commonly a forecast of the state by the underlying
model, initialized by the analysis state in the previous time step.
One of the most common approaches to the data assimilation problem
relies on variational techniques \citep[e.g.,][among many others]{[Sas58],[Sas70],[Lor86],[TalC87],[CouT90],[ParD92],[Zup93],[CouTH94],[ReiME01]}.
In these methods, one explicitly defines a cost function, typically
quadratic, whose potentially unique minimizer is the analysis state.
Very recently \citet{[FreNB12]} proposed a regularized data assimilation
scheme to improve assimilation results in advective sharp fluid fronts.

What is common in the DS, DF, and DA problems is that, in all of them,
we seek an improved estimate of the true state given a suite of noisy
and down-sampled observations and/or uncertain model-predicted states.
Specifically, let us suppose that the unknown \emph{true }state in
continuous space is denoted by $x(t)$ and its indirect observation
(or model-output), by $y(r)$. Let us also assume that $x(t)$ and
$y(r)$ are related via a linear integral equation, called the Fredholm
integral equation of the the first kind, as:
\begin{equation}
\int_{0}^{1}\mathcal{H}(r,\, t)\, x(t)\, dt=y(r),\,\,\,\,\,\,\,0\leq r\leq1,\label{eq:1}
\end{equation}
where $\mathcal{H}(r,\, t)$ is the kernel relating $x(t)$ and $y(r)$.
Recovery of $x(t)$ knowing $y(r)$ and $\mathcal{H}(r,\, t)$ is
a classic linear inverse problem. Clearly, the deconvolution problem
is a very special case with the kernel of the form $\mathcal{H}(r-t)$,
which in its discrete form, plays a central role in this paper. Linear
inverse problems are by nature ill-posed, in the sense that they do
not satisfy at least one of the following three conditions: (1) Existence;
(2) Uniqueness; and (3) Stability of the solution. For instance, when
due to the kernel architecture, the dimension of the observation is
smaller than that of the true signal, infinite choices of $x(t)$
lead to the same $y(r)$ and there is no unique solution for the problem.
For the case when $y(r)$ is noisy and has a larger dimension than
the true state, the solution is typically very unstable, because,
the high frequency components in $y(r)$ are typically amplified and
spoil the solution. In fact, the higher the frequency, the larger
the amplification in the solution, which is often called the inverted
noise; see, e.g., \citet{[Han10]} for a comprehensive account on
linear inverse problems. A common approach to make an inverse problem
well-posed is via the so-called \emph{regularization} methods. The
goal of regularization is to reformulate the inverse problem aiming
to obtain a unique and sufficiently stable solution. The choice of
regularization typically depends on the continuity and degree of smoothness
of the state variable of interest, often called the \emph{regularity
}condition. For instance, some state variables are very regular with
high degree of smoothness (e.g., pressure) while others might be more
irregular and suffer from frequent and different sorts of discontinuities
(e.g., rainfall). In fact, the choice of regularization not only yields
unique and stable solutions but also reinforces the underlying regularity
of the true state in the solution. It is important to note that, different
regularity conditions are theoretically consistent with different
statistical signatures in the true state, a fact that may guide proper
design of the regularization, as explored in this study.

The central goal of this paper is to propose a unified framework for
the class of DS, DF, and DA problems by recasting them as discrete
linear inverse problems using relevant regularizations in the derivative
space aiming to solve them more accurately compared to the classic
formulations. Examples on rainfall DS and DF are presented to quantitatively
elaborate on the effectiveness of the presented methodologies. In
these examples, regularization is performed consistent with the non-Gaussian
statistics of rainfall in the derivative space, which might be generalized
to other land-surface signals such as soil moisture. For the DA family
of problems, the promise of the presented framework, is demonstrated
via an elementary example using the heat equation. It turns out that
the accuracy of the analysis and forecast cycles can be improved compared
to the classic DA methods, especially when the initial state is discontinuous.
This simple example provides insight into how the new formulations
can outperform the classic methods and be of special interest in hydrometeorological
applications.

This paper is structured as follows. Section 2 provides conceptual
insight into the discrete inverse problems. Section 3 describes the
DS problem in detail, as a primitive building block for the other
studied problems. Important classes of regularization methods are
explained and their statistical interpretation is discussed from a
Bayesian point of view. Examples on rainfall downscaling are presented
in this section, taking into account the specific regularity and statistical
distribution of the rainfall fields in the derivative space. Section
3 is devoted to the regularized DF class of problems with examples
and results again on rainfall data. The regularized DA problem is
discussed in Section 4 with an illustrative example using the heat
equation. Concluding remarks and future research perspectives are
discussed in Section 5.

\section{Discrete Inverse problems: Conceptual Framework}

In this section, we briefly explain the conceptual key elements of
discrete linear inverse estimation relevant to the problems at hand
and leave further details for the next sections. Analogous to equation
(\ref{eq:1}), linear discrete inverse problems typically amount to
estimating the true $m$-element state vector $\mathbf{x}\in\mathbb{R}^{m}$
from the following linear observation model:
\begin{equation}
\mathbf{y}=\mathbf{H}\mathbf{x}+\mathbf{v},\label{eq:2}
\end{equation}
where $\mathbf{y}\in\mathbb{R}^{n}$ denotes the measurement, e.g.,
output of a sensor, $\mathbf{H}\in\mathbb{R}^{n\times m}$ is an $n$-by-$m$
observation matrix and $\mathbf{v}\sim\mathcal{N}\left(0,\,\mathbf{R}\right)$
is the Gaussian error in $\mathbb{R}^{n}$. Note that, the observation
operator, which is a discrete representation of the kernel in (\ref{eq:1}),
and the noise covariance are supposed to be known or properly calibrated.
Depending on the relative dimension of $\mathbf{y}$ and $\mathbf{x}$,
this linear system can be under-determined ($m\gg n$) or over-determined
($m\ll n$) (see, Figure \ref{Fig_1}). Provided that $\mathbf{H}$
is full rank, in the first case, there are infinite different $\mathbf{x}$'s
that satisfy (\ref{eq:2}) while for the second case a single exact
solution does not exist. As is evident, the DS class of problems belongs
to the under-determined class because the sensor output is a coarse-scale
and noisy representation of the true state. However, the class of
DF and DA problems fall into the category of over-determined problems,
as the total size of the observations and background state exceeds
the dimension of the true state (Figure \ref{Fig_1}).

\begin{figure}
\noindent \begin{centering}
$\begin{array}{c}
{\rm a)}\\
\\
\\
\\
\end{array}\,\,\begin{bmatrix}|\\
\mathbf{y}\\
|
\end{bmatrix}=\begin{bmatrix}\, & \, & \, & \, & \,\\
| & \cdots & \mathbf{H} & \cdots & |\\
\, & \, & \, & \, & \,
\end{bmatrix}\begin{bmatrix}|\\
\vdots\\
\mathbf{x}\\
\vdots\\
|
\end{bmatrix}+\begin{bmatrix}|\\
\mathbf{v}\\
|
\end{bmatrix}\,\,\,\,\,\,\,\,\,\,\,\,\,\,\,\,\,\,\,\,\,\,\,\,\,\,\,\,\,\,\begin{array}{c}
{\rm b)}\\
\\
\\
\\
\end{array}\,\,\begin{bmatrix}|\\
\vdots\\
\mathbf{y}\\
\vdots\\
|
\end{bmatrix}=\begin{bmatrix}\, & | & \,\\
\, & \vdots & \,\\
\, & \mathbf{H} & \,\\
\, & \vdots & \,\\
\, & | & \,
\end{bmatrix}\begin{bmatrix}|\\
\mathbf{x}\\
|
\end{bmatrix}+\begin{bmatrix}|\\
\vdots\\
\mathbf{v}\\
\vdots\\
|
\end{bmatrix}$
\par\end{centering}

\caption{Schematics of under-determined (a) and over-determined (b) linear
discrete observation model in (\ref{eq:2}). (a) The observation matrix
is fat with more columns than rows and $\mathbf{x}$ has a greater
dimension than $\mathbf{y}$. (b) The observation matrix is skinny
with more rows than columns and size of $\mathbf{y}$ exceeds the
dimension of $\mathbf{x}$. \label{Fig_1}}
\end{figure}

In each of the above cases, we may naturally try to obtain a solution
with minimum error variance by solving a linear least squares (LS)
problem. However, for the under-determined case the solution still
does not exist, while for the over-determined case it is commonly
ill-conditioned and sensitive to the observation noise (see, Section
4). Therefore, the minimum variance LS treatment can not properly
make the above inverse problems well-posed. To obtain a unique and
stable solution, beyond the LS minimization of the error, the basic
idea of regularization is to further constrain the solution. For instance,
among many solutions that fit the observation model in (\ref{eq:2}),
we can obtain the one with minimum energy, mean-squared curvature
or total variation. The choice of this constrain or regularization,
highly depends on a priori knowledge about the underlying regularity
of $\mathbf{x}$. For sufficiently smooth $\mathbf{x}$ we naturally
may promote a solution with minimum mean-squared curvature to impose
smoothness on the solution. However, if a state is non-smooth and
contains frequent jump discontinuities, a solution with minimum total
variation might be a better choice. In subsequent sections, we explain
these concepts in more detail for the DS, DF, and DA problems with
hydrometeorological examples.

\section{Regularized Downscaling }

\subsection{Problem Formulation}

To put the DS problem in a linear inverse estimation framework, we
recognize that in the observation model of equaiton (\ref{eq:2}),
the true state $\mathbf{x}\in\mathbb{R}^{m}$ has a larger dimension
than the observation vector $\mathbf{y}\in\mathbb{R}^{n}$, $m\gg n$.
Note that throughout this work, a notation is adopted in which the
vector \textbf{$\mathbf{x}\in\mathbb{R}^{m}$} may also represent,
for example a 2D field \textbf{${\rm X}\in\mathbb{R}^{\sqrt{m}\times\sqrt{m}}$}
which is vectorized in a fixed order (e.g., lexicographical).

As explained in the previous section, the DS problem naturally amounts
to obtaining the best weighted least squares (WLS) estimate $\mathbf{\hat{x}}$
of the high-resolution or fine-scale true state as
\begin{equation}
\hat{\mathbf{x}}=\underset{\mathbf{x}}{{\rm argmin}}\left\{ \frac{1}{2}\left\Vert \mathbf{y}-\mathbf{Hx}\right\Vert _{\mathbf{R}^{-1}}^{2}\right\} ,\label{eq:3}
\end{equation}
where, $\left\Vert \mathbf{x}\right\Vert _{\mathbf{A}}^{2}=\mathbf{x}^{T}\mathbf{A}\mathbf{x}$
denotes the \emph{quadratic-norm,} while $\mathbf{A}$ is a positive
definite matrix. In the above notation, $\frac{1}{2}\left\Vert \mathbf{y}-\mathbf{Hx}\right\Vert _{\mathbf{R}^{-1}}^{2}$
is called the cost function whose minimum is attained at $\hat{\mathbf{x}}$.
Due to the ill-posed nature of the problem, this optimization does
not have a unique solution. Specifically, taking the derivative and
setting it to zero we get
\begin{equation}
\left(\mathbf{H}^{T}\mathbf{R}^{-1}\mathbf{H}\right)\hat{\mathbf{x}}=\mathbf{H}^{T}\mathbf{R}^{-1}\mathbf{y},
\end{equation}
in which $\mathbf{H}^{T}\mathbf{R}^{-1}\mathbf{H}$ is definitely
singular. Indeed, many choices of $\hat{\mathbf{x}}$ lead to the
same right-hand side. To narrow down all possible solutions to a stable
and unique one, a common choice is to regularize the problem by constraining
the squared Euclidean norm of the solution to be less than a certain
constant, that is $\left\Vert \mathbf{L}\mathbf{x}\right\Vert _{2}^{2}\leq{\rm const.}$,
where $\mathbf{L}$ is an appropriately chosen transformation and
$\left\Vert \mathbf{x}\right\Vert _{2}^{2}=\sum_{i}x_{i}^{2}$ denotes
the Euclidean $l_{2}$-norm. Note that, by putting a constraint on
the Euclidean norm of the state, we not only narrow down the solutions
but also implicitly suppress the large components of the inverted
noise and reduce their spoiling effect on the solution.

Using the theory of Lagrangian multipliers the dual form of the constrained
version of the optimization in (\ref{eq:3}) is
\begin{equation}
\hat{\mathbf{x}}=\underset{\mathbf{x}}{{\rm argmin}}\left\{ \frac{1}{2}\left\Vert \mathbf{y}-\mathbf{Hx}\right\Vert _{\mathbf{R}^{-1}}^{2}+\lambda\,\left\Vert \mathbf{L}\mathbf{x}\right\Vert _{2}^{2}\right\} ,\label{eq:5}
\end{equation}
where $\lambda>0$ is the Lagrangian multiplier or the so-called \emph{regularizer}.

The first term in (\ref{eq:5}) measures how well the solution approximates
the given (noisy) data, while the second term imposes a specific regularity
on the solution. In effect, the regularizer plays a trade off role
between making the fidelity to the observations sufficiently large,
while not imposing too much regularity (in this case, smoothness)
on the solution. The smaller the value of $\lambda$, the more weight
is given to fitting the (noisy) observations resulting in solutions
that are less regular and prone to overfitting. On the other hand,
the larger the value of $\lambda$ the more weight is given to the
regularity of the solution. The goal is to find a regularized solution,
by finding a good balance between the two terms such that the it is
sufficiently close to the observations while obeying the underlying
property of the true state. As is evident, the $\mathbf{L}$-transformation
of the state also plays a key role in the solution of equation (\ref{eq:5}).
For instance, choosing an identity matrix implies that we are looking
for a solution with a small Euclidean norm (energy), the so called
\emph{least-norm solution}, while if $\mathbf{L}$ represents an approximation
of the second order derivative (i.e., Laplacian transform), the $\left\Vert \mathbf{L}\mathbf{x}\right\Vert _{2}^{2}$
is a measure of the mean-square curvature of the state which imposes
extra smoothness on the solution.

The problem in equation (\ref{eq:5}) is a \emph{smooth} convex quadratic
programming problem since the cost function is differentiable and
its Hessian $\mathbf{H}^{T}\mathbf{R}^{-1}\mathbf{H}+2\lambda\mathbf{L}^{T}\mathbf{L}$
is always positive definite for any $\lambda>0$, provided that $\mathbf{L}^{T}\mathbf{L}$
is positive definite. This problem is known as the \emph{Tikhonov
regularization} with the following analytical solution
\begin{equation}
\hat{\mathbf{x}}=\left(\mathbf{H}^{T}\mathbf{R}^{-1}\mathbf{H}+2\lambda\,\mathbf{L}^{T}\mathbf{L}\right)^{-1}\mathbf{H}^{T}\mathbf{R}^{-1}\mathbf{y},\label{eq:6}
\end{equation}
\citep{[Tik77]}. It is easy to show that the covariance of the estimate
is the inverse of the Hessian of the cost function in (\ref{eq:5})
which is $\left(\mathbf{H}^{T}\mathbf{R}^{-1}\mathbf{H}+2\lambda\,\mathbf{L}^{T}\mathbf{L}\right)^{-1}$.
Notice that, for large scale regularization problems, this closed
form solution can not be directly computed and efficient iterative
methods (e.g., Newton's method) need to be employed.

Depending on the selected transformation and the intrinsic regularity
(degree of smoothness) of the underlying state, other choices of the
regularization term are also common. For example, in case the $\mathbf{L}$-transformation
projects the major body of the state vector onto (near) zero values,
a common choice is the $l_{1}$-norm instead of the $l_{2}$-norm
in (\ref{eq:5}) \citep[e.g.,][]{[CheDS98]}. Such a property is often
referred to as \emph{sparse representation} in the $\mathbf{L}$-transformation
space and gives rise to the following formulation of the regularized
DS problem:
\begin{equation}
\hat{\mathbf{x}}=\underset{\mathbf{x}}{{\rm argmin}}\left\{ \frac{1}{2}\left\Vert \mathbf{y}-\mathbf{Hx}\right\Vert _{\mathbf{R}^{-1}}^{2}+\lambda\,\left\Vert \mathbf{L}\mathbf{x}\right\Vert _{1}\right\} ,\label{eq:7}
\end{equation}
where, the $l_{1}$-norm is $\left\Vert \mathbf{x}\right\Vert _{1}=\sum_{i}\left|x_{i}\right|$.
Note that, the problem in (\ref{eq:7}) is a \emph{non-smooth} convex
optimization as the regularization term is non-differentiable and
the conventional iterative gradient descent methods are no longer
applicable in their standard forms. Several optimization techniques
are currently under development to directly solve the non-smooth inverse
problem in (\ref{eq:7}) that may include: (1) The iterative shrinkage
thresholding algorithms pioneered by \citet{[Tib96]}; (2) The basis
pursuit by \citet{[CheDS98]}; (3) Constrained quadratic programing
\citep[e.g.,][]{[FigNW07]}; (4) Proximal gradient based methods \citep[e.g.,][]{[BecT09a]};
and (5) Interior point methods \citep[e.g.,][]{[KimKLB07]}.

\begin{figure}[H]
\noindent \begin{centering}
\includegraphics[scale=0.5]{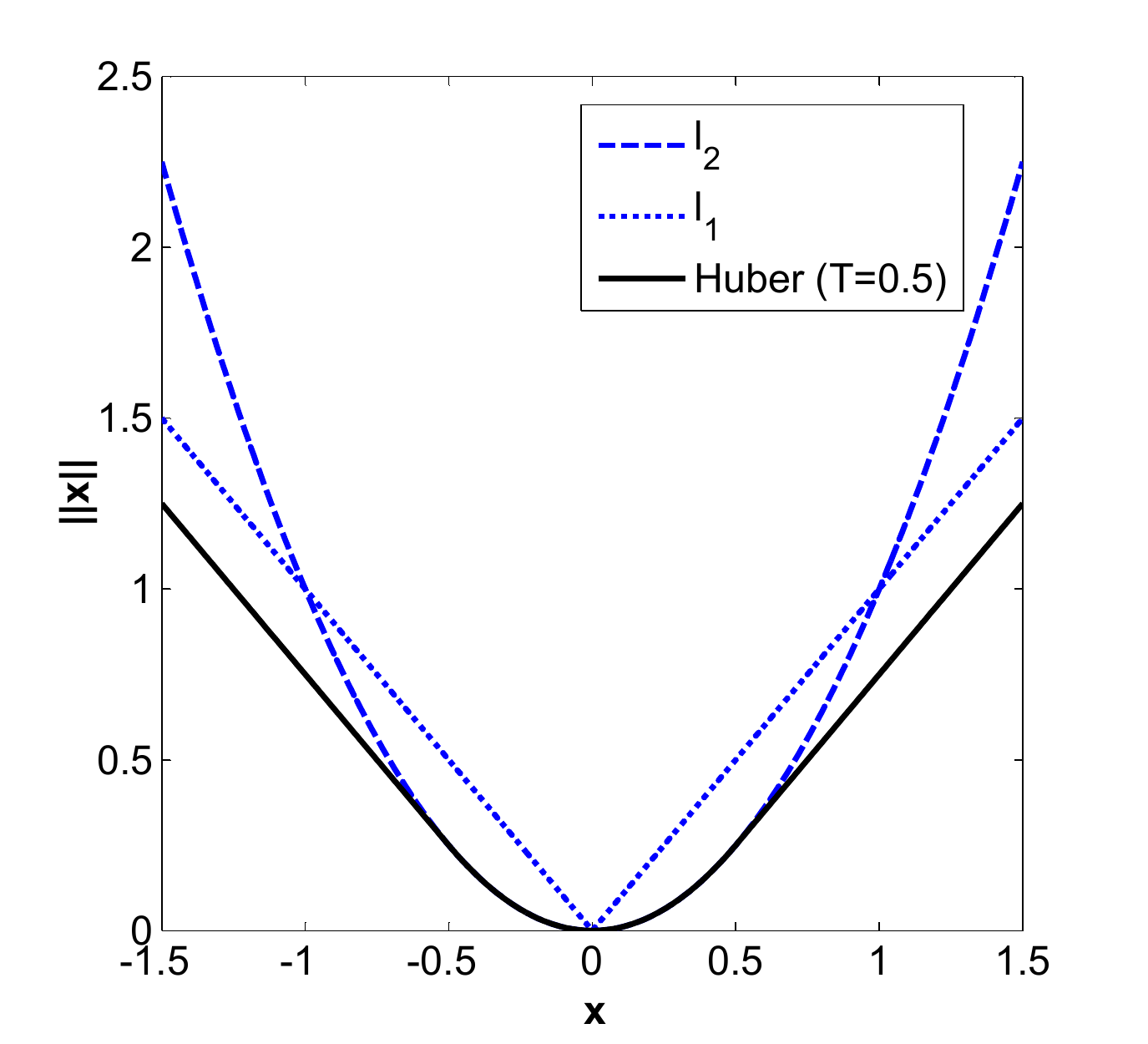}
\par\end{centering}

\caption{The $l_{1}$, $l_{2}$ and Huber norms. The Huber penalty is a smooth
relaxation of the $l_{1}$-norm which acts quadratically for input
values smaller than the threshold while it behaves linearly for larger
inputs. For inputs with heavy tail behavior, linear penalization in
the regularization term is advantageous compared to the quadratic
penalty functions in which the cost function becomes dominated by
a few large values in the tail of the distribution. \label{Fig_2}}
\end{figure}

One of the common approaches to treat the non-differentiability in
(\ref{eq:7}) is to replace the $l_{1}$-norm with a smooth approximation,
the so called Huber norm, $\left\Vert \mathbf{x}\right\Vert _{{\rm Hub}}=\sum_{i}\rho_{T}(x_{i})$,
where
\begin{equation}
\rho_{T}(x)=\begin{cases}
x^{2} & \left|x\right|\leq T\\
T\left(2\left|x\right|-T\right) & \left|x\right|>T
\end{cases},
\end{equation}

and $T$ denotes a non-negative threshold (Figure \ref{Fig_2}). The
Huber norm is a hybrid one that behaves similar to the $l_{1}$-norm
for values greater than the threshold $T$ while for smaller values
it is identical to the $l_{2}$-norm. From the statistical regression
point of view, the sensitivity of a norm as a penalty function to
the outliers depends on the (relative) values of the norm for large
residuals. If we restrict ourselves to convex norms, the least sensitive
ones to the large residuals or say the outliers are those with linear
behavior for large input arguments (i.e., $l_{1}$ and Huber). Because
of this property, these $l_{1}$-like norms are often called \emph{robust}
norms, \citep{[hub64],[Hub81],[BoyV04]}. Throughout this paper, for
solving $l_{1}$-regularized inverse problems, we use the Huber relaxation
due to its simplicity, efficiency and adaptivity to all of the concerning
classes of DS, DF, and DA problems.

In downscaling of state variables containing frequent jumps and isolated
singularities (e.g., local rainfall extremes) using regularization
in the derivative space (i.e., $\mathbf{L}\mathbf{x}$ is an approximate
derivative measure), the $l_{1}$-like norms typically preserve rapid
variations and lead to improved and sharper solutions, compared to
the overly smooth results of the $l_{2}$-regularization; see, examples
presented by \citet{[EbtF12a]} for rainfall downscaling and other
examples in the next section.

\subsection{Statistical Interpretation}

From the \emph{frequentist} statistical point of view, it is easy
to show that the WLS solution of (\ref{eq:3}) is equivalent to the
maximum likelihood estimator (ML)
\begin{equation}
\hat{\mathbf{x}}_{ML}=\underset{\mathbf{x}}{{\rm argmax}}\, p\left(\mathbf{y}|\mathbf{x}\right),
\end{equation}
given that the likelihood density is Gaussian, $p\left(\mathbf{y}|\mathbf{x}\right)\propto\exp\left(\nicefrac{-1}{2}(\mathbf{y}-\mathbf{H}\mathbf{x})^{T}\mathbf{R}^{-1}(\mathbf{y}-\mathbf{Hx})\right)$.
Specifically, taking $-\log(\cdot)$, one can find the minimizer of
the negative log-likelihood function $-\log\left\{ p(\mathbf{y}|\mathbf{x})\right\} $
as,
\begin{eqnarray}
\hat{\mathbf{x}}_{ML} & = & \underset{\mathbf{x}}{{\rm argmin}}\left\{ \frac{1}{2}(\mathbf{y}-\mathbf{H}\mathbf{x})^{T}\mathbf{R}^{-1}(\mathbf{y}-\mathbf{Hx})\right\} \nonumber \\
 & = & \underset{\mathbf{x}}{{\rm argmin}}\left\{ \frac{1}{2}\left\Vert \mathbf{y}-\mathbf{Hx}\right\Vert _{\mathbf{R}^{-1}}^{2}\right\} ,
\end{eqnarray}
which is identical to the WLS solution of (\ref{eq:3}).

It is important to note that in the ML estimator, $\mathbf{x}$ is
considered to be a deterministic variable (fixed) while $\mathbf{y}$
has a random nature.

On the other hand, in the \emph{Bayesian} perspective, the regularized
solution of equation (\ref{eq:5}) is equivalent to the maximum a
posteriori (MAP) estimator
\begin{equation}
\hat{\mathbf{x}}_{MAP}=\underset{\mathbf{x}}{{\rm argmax}}\, p\left(\mathbf{x}|\mathbf{y}\right),
\end{equation}
where both $\mathbf{x}$ and $\mathbf{y}$ are considered of random
nature. Specifically, using Bayes theorem, ignoring the constant terms
in \textbf{$\mathbf{x}$ }and applying $-\log(\cdot)$ on the posterior
density $p(\mathbf{x}|\mathbf{y})$, we get
\begin{eqnarray}
\hat{\mathbf{x}}_{MAP} & = & \underset{\mathbf{x}}{{\rm argmin}}\left\{ -\log\,\left(\frac{p(\mathbf{y}|\mathbf{x})\, p(\mathbf{x})}{p(\mathbf{y})}\right)\right\} \nonumber \\
 & = & \underset{\mathbf{x}}{{\rm argmin}}\left\{ -\log\, p\left(\mathbf{y}|\mathbf{x}\right)-\log\, p(\mathbf{x})\right\} .\label{eq:12}
\end{eqnarray}
The first term, $-\log\, p\left(\mathbf{y}|\mathbf{x}\right)$, is
just the negative log-likelihood as appeared in the ML estimator and
the second term is called the \emph{prior} which accounts for the
a priori knowledge about the density of the state vector $\mathbf{x}$.
Accordingly, the proposed Tikhonov regularization in (\ref{eq:5})
is equivalent to the MAP estimator assuming that the state, or the
linear transformed state $\mathbf{L}\mathbf{x}$, is a multivariate
Gaussian of the form
\begin{equation}
\log\, p(\mathbf{x})\propto\mathbf{x}^{T}\mathbf{Q}\mathbf{x},
\end{equation}
where the covariance is $\mathbf{Q}=\mathbf{L}^{T}\mathbf{L}$ \citep[e.g.,][]{[Tik77],[ElaF97],[Lev08]}.

Clearly, the choice of the $l_{1}$-norm in equation (\ref{eq:7}),
implies that $\log\, p(\mathbf{x})\propto\left\Vert \mathbf{L}\mathbf{x}\right\Vert _{1}$
or say the transformed state can be well explained by a multivariate
\emph{Laplace }density with heavier tail than the Gaussian case \citep[e.g.,][]{[Tib96],[LewS00]}.
Similarly, selecting the Huber norm can also be interpreted as $\log\, p(\mathbf{x})\propto\sum_{i}\rho_{T}(x_{i})$,
which is equivalent to assuming the \emph{Gibbs }density function
as the prior probability model of the state \citep{[GemG84],[SchS94]}.
The equivalence between the regularized estimation approach, which
imposes constraints on the regularity of the solution, and its Bayesian
counterpart, which imposes constraints on the prior probability of
the state is very insightful. This relationship establishes an important
duality which can guide the selection of regularization depending
on the statistical properties of the state in the real or derivative
space.

\subsection{Examples on Rainfall DS}

\subsubsection{Problem Formulation and Settings}

\begin{figure}[t]
\noindent \begin{centering}
${\rm a)}$
\par\end{centering}

\noindent \begin{centering}
\includegraphics{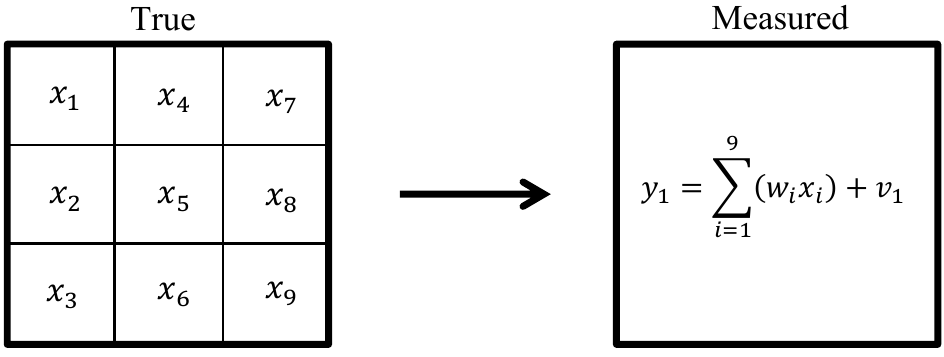}
\par\end{centering}

\medskip{}

\noindent \begin{centering}
$\overset{{\rm Kernel}}{\begin{bmatrix}c_{1} & c_{4} & c_{7}\\
c_{2} & c_{5} & c_{8}\\
c_{3} & c_{6} & c_{9}
\end{bmatrix}}\,\,\,\,\,\overset{{\rm b)\, Filtering}}{\begin{array}{c}
\hrulefill\\
\begin{array}{ccccc}
0 & 0 & 0 & 0 & 0\\
0 & 0 & 0 & 0 & 0\\
0 & 0 & 1 & 0 & 0\\
0 & 0 & 0 & 0 & 0\\
0 & 0 & 0 & 0 & 0
\end{array}\xrightarrow{\mathbf{C}}\begin{array}{ccccc}
0 & 0 & 0 & 0 & 0\\
0 & c_{9} & c_{6} & c_{3} & 0\\
0 & c_{8} & c_{5} & c_{2} & 0\\
0 & c_{7} & c_{4} & c_{1} & 0\\
0 & 0 & 0 & 0 & 0
\end{array}
\end{array}}\,\,\,\,\,\,\,\,\,\overset{{\rm c)\, Convolution}}{\begin{array}{c}
\hrulefill\\
\begin{array}{ccccc}
0 & 0 & 0 & 0 & 0\\
0 & 0 & 0 & 0 & 0\\
0 & 0 & 1 & 0 & 0\\
0 & 0 & 0 & 0 & 0\\
0 & 0 & 0 & 0 & 0
\end{array}\xrightarrow{\mathbf{C}^{T}}\begin{array}{ccccc}
0 & 0 & 0 & 0 & 0\\
0 & c_{1} & c_{4} & c_{7} & 0\\
0 & c_{2} & c_{5} & c_{8} & 0\\
0 & c_{3} & c_{6} & c_{9} & 0\\
0 & 0 & 0 & 0 & 0
\end{array}
\end{array}}$
\par\end{centering}

\bigskip{}

\noindent \begin{centering}
$\,\overset{{\rm d)\, Upsampling\, and}\,{\rm Downsampling}}{\begin{array}{c}
\hrulefill\\
\begin{array}{ccc}
1 & 1 & 1\\
1 & 1 & 1\\
1 & 1 & 1
\end{array}\begin{array}{c}
\xrightarrow{\mathbf{D}^{T}}\\
\xleftarrow[\mathbf{D}]{}
\end{array}\begin{array}{ccccc}
1 & 0 & 1 & 0 & 1\\
0 & 0 & 0 & 0 & 0\\
1 & 0 & 1 & 0 & 1\\
0 & 0 & 0 & 0 & 0\\
1 & 0 & 1 & 0 & 1
\end{array}
\end{array}}$
\par\end{centering}

\caption{Two dimensional mathematical models of the smoothing and downsampling
property of a low-resolution sensor via the convolution operation.
(a) A simple representation of the used observation model for a neighborhood
of size 3-by-3 using a simple uniform averaging filter. (b-c) A sample
effect of the filtering operation ($\mathbf{C}$) and its transpose
($\mathbf{C}^{T}$) on a discrete 2D unit pulse, given the shown $3\times3$
kernel. (d) A sample effect of the 2D downsampling operator ($\mathbf{D}$)
and its transpose ($\mathbf{D}^{T}$) with scaling ratio 2. In the
filtering operation, we basically slide the kernel over the field
and sum the element-wise multiplication of the kernel elements with
those of the field and then put the results at the center of the current
position of the kernel. However, in the convolution operation we first
rotate the kernel by $180^{\circ}$ and follow the same procedure.\label{Fig_3}}
\end{figure}

To downscale a remotely sensed hydrometeorological signal using the
explained discrete regularization methods, we need to have proper
mathematical models for the downgrading operator and also a priori
knowledge about the form of the regularization term. Clearly, the
downgrading operator in the presented framework needs to be a linear
approximation of the sampling property of the sensor. If a sensor
directly measures the state of interest, while its maximum frequency
channel is smaller than the maximum frequency content of the state
(e.g., precipitation), the result of the sensing would be a smoothed
and possibly downsampled version of the true state. Thus, each element
of the observed state in a grid-scale might be considered as an average
(low-resolution) representation of the true state, lacking the high-resolution
subgrid-scale variability. To have a simple and tractable mathematical
model, the downgrading matrix might be considered translation invariant
and decomposed into $\mathbf{H}=\mathbf{D}\mathbf{C}$, where $\mathbf{C}$
encodes the smoothing effect and $\mathbf{D}$ contains information
about the sampling rate of the sensor. As a simple mathematical model,
let us suppose that each grid point in the low-resolution observation
is a (weighted) average of a finite size neighborhood of the true
state, positioned at the center of the grid. In this case, the sensor
smoothing property in $\mathbf{C}$ can be encoded by the filtering
and convolution operations, while $\mathbf{D}$ acts as a linear downsampling
operator (Figure \ref{Fig_3}). These matrices can be formed explicitly,
while direct matrix-vector multiplication (e.g., $\mathbf{C}\mathbf{x}$
and $\mathbf{C}^{T}\mathbf{x}$, $\mathbf{x}\in\mathbb{R}^{m}$),
requires a computational cost in the order of $\mathcal{O}(m^{2})$.
However, for large scale problems, those matrix-vector multiplications
(i.e., convolution operation) are typically performed more efficiently,
for instance in the Fourier domain with computational cost of $\mathcal{O}(m\log m)$.
Figure \ref{Fig_3} sketches the above matrix-vector multiplication
via the filtering and convolution operations.

As is evident, the smoothing kernel needs to be estimated for each
sensor, possibly by learning from a library of coincidental high and
low-resolution observations (\citealp[e.g.,][]{[EbtF12a]}), or through
a direct minimization of an associated cost. In the absence of prior
knowledge, one possible choice is to assume that the sensor observes
a coarse grained (i.e., non-overlapping box averaging) and noisy version
of the true state. In other words, to produce a field at the grid-scale
of $s_{c}\times s_{c}$ from a $1\times1$, this assumption is equivalent
to selecting a uniform smoothing kernel of size $s_{c}\times s_{c}$
followed by a downsampling ratio of $s_{c}$ (Figure \ref{Fig_4}a).

\begin{figure}[H]
\noindent \begin{centering}
$\begin{array}{c}
{\rm a)}\\
\\
\\
\\
\end{array}\frac{1}{s_{c}^{2}}\begin{bmatrix}1 & \cdots & 1\\
\vdots & \ddots & \vdots\\
1 & \cdots & 1
\end{bmatrix}_{s_{c}\times s_{c}}\,\,\,\,\,\,\,\,\,\,\begin{array}{c}
{\rm b)}\\
\\
\\
\\
\end{array}\triangledown^{2}=\frac{4}{\left(\kappa+1\right)}\begin{bmatrix}\frac{\kappa}{4} & \frac{1-\kappa}{4} & \frac{\kappa}{4}\\
\frac{1-\kappa}{4} & -1 & \frac{1-\kappa}{4}\\
\frac{\kappa}{4} & \frac{1-\kappa}{4} & \frac{\kappa}{4}
\end{bmatrix}$
\par\end{centering}

\caption{(a) A uniform smoothing (low-pass) kernel of size $s_{c}\times s_{c}$.
(b) The discrete (high-pass) generalized Laplacian filter of size
$3\times3$, where $\kappa$ is a parameter ranging between 0 to 1.
The Laplacian coefficients, obtained by filtering the 2D state with
the Laplacian kernel, are approximate measures of the second order
derivative. Throughout this paper, we choose $\kappa=0.5$ which corresponds
to the 2D standard second order differencing operation.\label{Fig_4}}
\end{figure}

\begin{figure}[t]
\noindent \begin{centering}
\includegraphics[scale=0.45]{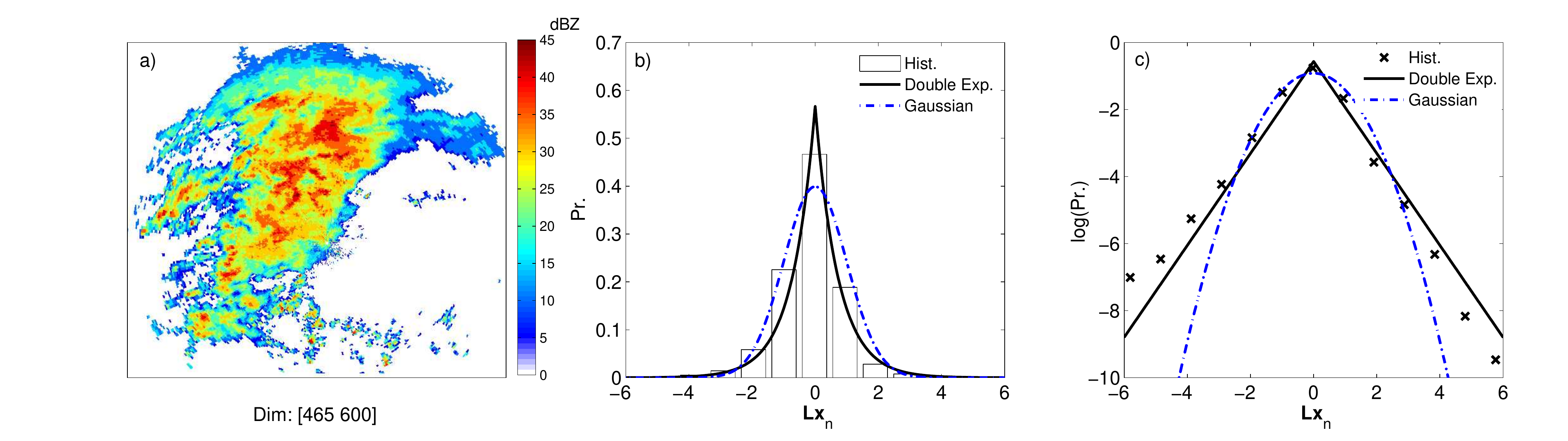}
\par\end{centering}

\caption{A rainfall reflectivity field and the distribution of its standardized
Laplacian coefficients with unit standard deviation, $\mathbf{L}\mathbf{x}_{{\rm n}}=\unitfrac{\mathbf{L}\mathbf{x}}{std(\mathbf{L}\mathbf{x})}$,
where ${\rm std(\cdot)}$ is the standard deviation operator and $\triangledown^{2}\mathbf{x}=\mathbf{L}\mathbf{x}$.
(a) NEXRAD reflectivity snapshot at the TRMM GV-site in Houston, TX
(HSTN) on 1998/11/13 (00:02:00 UTC) at scale $1\times1$ km. (b) The
histogram of the standardized Laplacian coefficients , with $\kappa=0.5$
(Figure \ref{Fig_4}), and (c) the log-histogram. Note that, the zero
coefficients over the non-rainy background have been excluded from
the histogram analysis. The dash line in (b) is the least squares
fitted exponential of the form $p(x)\propto\exp\left(-\lambda\left|x\right|\right)$
and the dash-dot line shows a standard normal distribution for comparison.
The log-histogram in (c) contrasts the heavy tailed structure of the
coefficients versus the Gaussian distribution clearer than the original
histogram in (b). \label{Fig_5}}
\end{figure}

The choice of the regularization term also plays a very important
role on the accuracy of the DS solution. Figure \ref{Fig_5}a demonstrates
a NEXRAD reflectivity snapshot ($1\times1$ km) over the Texas TRMM
ground validation (GV) site, while Figure \ref{Fig_5}b displays the
standardized histogram of the discrete Laplacian coefficients (second
order differences) and the fitted exponential of the form $p(x)\propto\exp\left(-\lambda\left|x\right|\right)$.
It is seen that the analyzed rainfall image exhibits a (near) sparse
representation in the derivative space with a large mass at zero and
heavier tail than the Gaussian. Although not shown here, the universality
of this structure can be observed in other rainfall reflectivity fields,
denoting that the choice of the $l_{1}$-like regularization is preferred
in the rainfall DS problems rather than the choice of the Tikhonov
regularization; see, \citet{[EbtF11a]} for a thorough survey of rainfall
statistics in derivative space, in terms of the wavelet coefficients.

This well behaved non-Gaussian structure in the derivative space mainly
arises due to the presence of spatial coherence (correlation) in the
rainfall fields and abrupt occurrences of piece-wise discontinuities
(large gradients). In effect, over the large areas of uniform rainfall
intensity, a measure of derivative translates rainfall values into
a large number of (near) zero values; however, over the less frequent
jumps and isolated high-intensity rain-cells, values of the derivative
measure are markedly larger than zero and form the tails. Note that
this non-Gaussianity is due to the intrinsic spatial structure of
rainfall reflectivity fields and can not be resolved by a logarithmic
or power-law transformation (e.g., $Z$-$R$ relationship). It is
easy to see that after applying the $Z$-$R$ relationship on the
reflectivity fields, the shape of the rainfall histogram remains non-Gaussian
and still can be explained by the Laplace density (not shown here).
In practice, the histogram of the derivatives may exhibit a thicker
tail than the Laplace density, requiring a heavier tail model such
as the Generalized Gaussian Density (GGD) of the form $p(x)\propto\exp\left(-\lambda\left|x\right|^{p}\right)$,
where $p<1$ \citep[see,][]{[EbtF11a]}. However, using such a prior
model gives rise to a non-convex optimization problem in which convergence
to the global minimum can not be guaranteed. Hence, the choice of
the $l_{1}$-norm (the Laplace prior) is indeed the closest convex
regularization that can partially fulfill the strict statistical interpretation,
while the actual prior might still be better explained by a heavier
tail model than the Laplace density.

Following our observations related to the distribution of the rainfall
derivatives, here we direct our attention to the Huber penalty function
as a smooth approximation of the $l_{1}$-regularization,

\begin{equation}
\mathcal{J}(\mathbf{x})=\frac{1}{2}\left\Vert \mathbf{y}-\mathbf{Hx}\right\Vert _{\mathbf{R}^{-1}}^{2}+\lambda\,\left\Vert \mathbf{L}\mathbf{x}\right\Vert _{{\rm Hub}}.\label{eq:14}
\end{equation}

Minimization of the above cost function can be easily achieved using
first order efficient gradient descent based methods. However, as
the rainfall fields are non-negative, we used the gradient projection
(GP) method \citep[pp. 228]{[Ber99]}, to solve the above problem
in a constrained mode such that $\mathbf{x}\succeq0$ (see Appendix
A).

\subsubsection{Results}

The same rainfall snapshot shown in Figure \ref{Fig_5} has been used
to examine the performance of the proposed regularized DS methodologies.
Throughout the paper, to make the reported parameters independent
of the range of intensity values, the rainfall reflectivity fields
are first scaled into the range between 0 and 1; however, the results
and Figures are presented in the true range.

To demonstrate the performance of the proposed regularized DS methodology,
the NEXRAD high-resolution observation $\mathbf{x}$ was assumed as
the true state while the low-resolution observations $\mathbf{y}$
were obtained by smoothing $\mathbf{x}$ with an average filter of
size $s_{c}\times s_{c}$ followed by a downsampling ratio $s_{c}$.
Given the true state and constructed observations, we can quantitatively
examine the effectiveness of the presented DS methods. In fact, selecting
some common quality metrics, we demonstrate the effective improvement
of those measures using the presented regularized DS framework.

\begin{figure}
\noindent \begin{centering}
\includegraphics{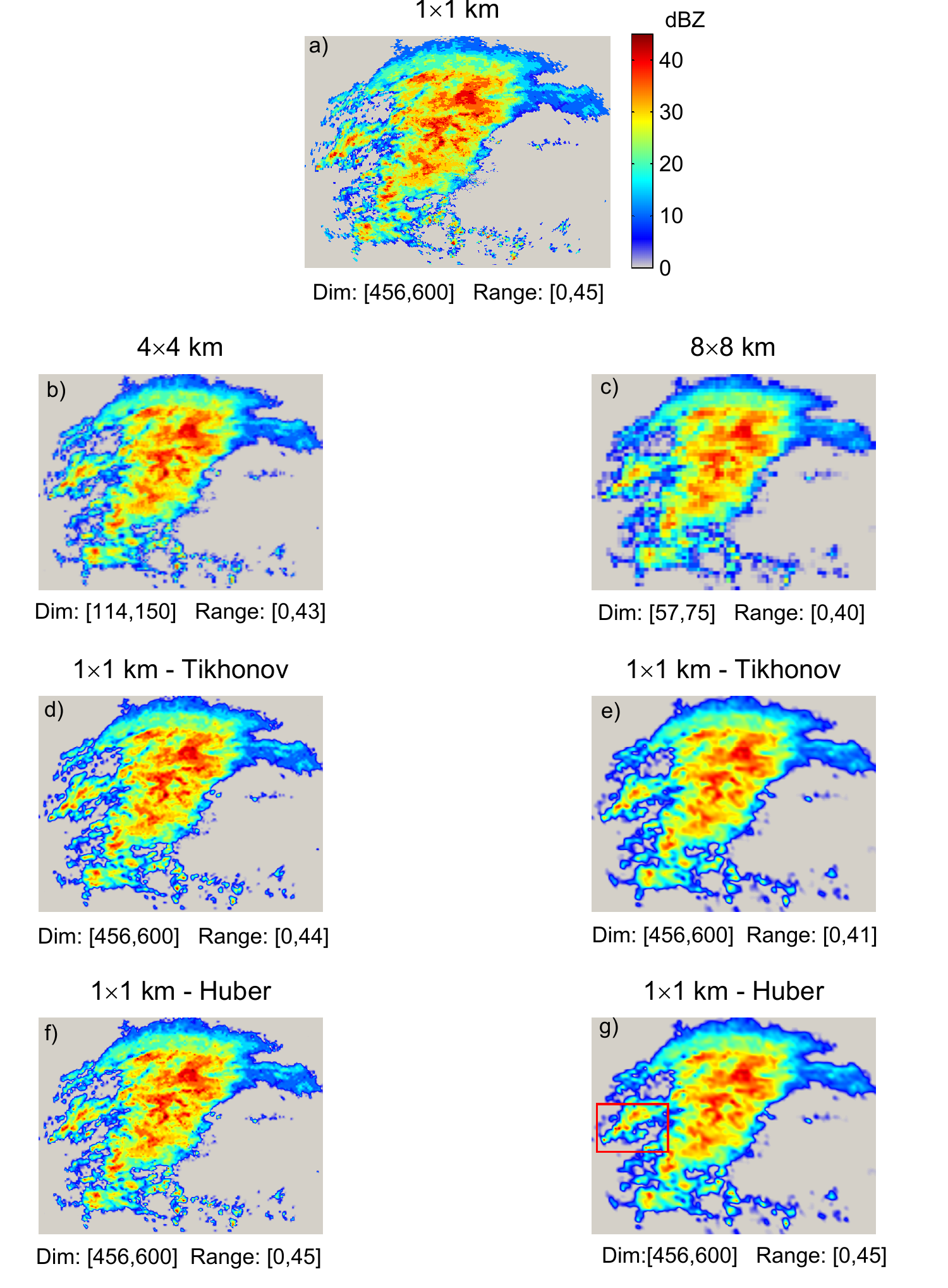}
\par\end{centering}

\caption{Sample results of the rainfall regularized Downscaling (DS). (a) True
high-resolution rainfall reflectivity: NEXRAD snapshot at the TRMM
GV-site in Houston, TX (HSTN) on 1998/11/13 (00:02:00 UTC) at resolution
$1\times1$ km. (b-c) The synthetically generated, $4\times4$ and
$8\times8$ km, coarse-scale and noisy observations of the true rainfall
reflectivity field. Left column: Tikhonov (d) and Huber (f) regularization
results for downscaling from 4-to-1 km ($T=0.02$). Right column:
Tikhonov (e) and Huber (g) regularized DS for downscaling from 8-to-1
km ($T=0.04)$. Zooming views of the delineated box in (g) are shown
in Figure \ref{Fig_7}.\label{Fig_6}}
\end{figure}

\begin{figure}[t]
\noindent \begin{centering}
\includegraphics[scale=0.55]{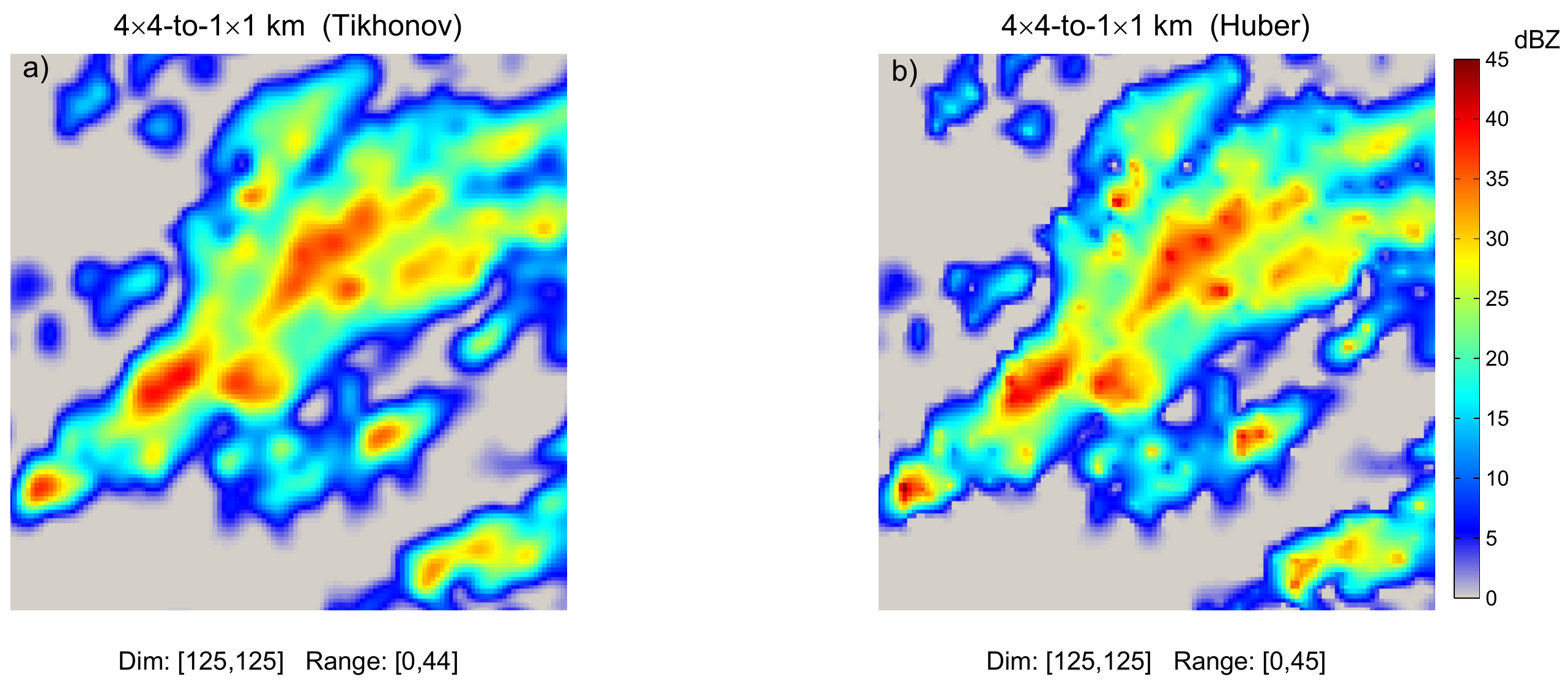}
\par\end{centering}

\noindent \begin{centering}
\includegraphics[scale=0.55]{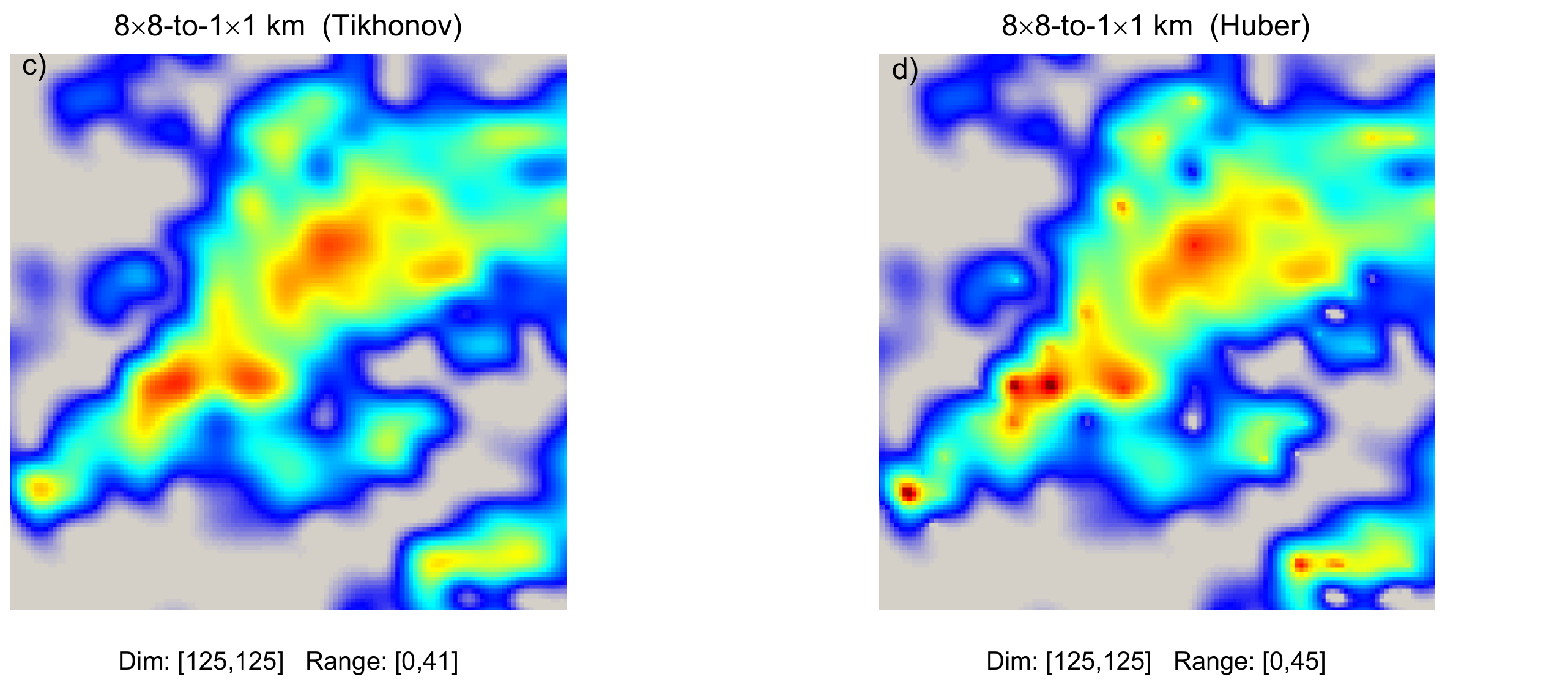}
\par\end{centering}

\caption{A zooming view for comparing qualitatively the Tikhonov (a, c) versus
Huber (b, d) regularization, for the Downscaling (DS) example over
the delineated box in Figure \ref{Fig_6}g. The results indicate a
better performance of the Huber regularization, especially for smaller
scaling ratios. The Huber regularization yields sharper results and
is more capable to recover high-intensity rainfall cells and the correct
range of variability; see Table \ref{Tab_1} for quantitative comparison
using a suit of metrics.\label{Fig_7}}
\end{figure}

Both the Huber and Tikhonov regularizations were examined to downscale
the observations from scales $4\times4$ and $8\times8$ km down to
$1\times1$ km (Figure \ref{Fig_6}). A very small amount of white
noise (i.e., standard deviation of $1{\rm e}$-3) was added to the
low-resolution observations (equation \ref{eq:2}), giving rise to
a diagonal error covariance matrix. In both of the regularization
methods, the regularization parameter $\lambda$ was set to $5{\rm e}$-3
and ${\rm 1e}$-2 for downscaling from 4-to-1 and 8-to-1 km in grid
spacing, respectively. These values are selected through trial and
error; however, there are some formal methods for automatic estimation
of this parameter which are left for future work \citep[e.g., the L-curve; see][]{[Han10]}.
In our experiments, it turned out that small values of the Huber threshold
$T$, typically less than $10$\% of the field maximum range of variability,
lead to a successful recovery of isolated singularities and local
extreme rainfall cells (Figure \ref{Fig_7}).

In the studied snapshot, coarse graining of the rainfall reflectivity
fields to the scales of $4\times4$ and $8\times8$ kilometers was
equivalent to loosing almost 20 and 30 percent of the rainfall energy
in terms of the relative Root Mean Squared Error (RMSE), ${\rm RMSE}_{{\rm r}}=\nicefrac{\left\Vert \mathbf{x}-\hat{\mathbf{x}}\right\Vert _{2}}{\left\Vert \mathbf{x}\right\Vert _{2}}$
(see, Table \ref{Tab_1}). Note that, to compute the RMSE of the low-resolution
observations, the size of those fields was extended to the size of
the true field using the nearest neighborhood interpolation, that
is, each low-resolution pixel was replaced with $s_{c}\times s_{c}$
pixels with the same intensity value. In addition to the relative
RMSE measure, we also used three other metrics: (1) Relative Mean
Absolute Error (MAE), ${\rm MAE}_{{\rm r}}=\nicefrac{\left\Vert \mathbf{x}-\hat{\mathbf{x}}\right\Vert _{1}}{\left\Vert \mathbf{x}\right\Vert _{1}}$;
(2) A logarithmic measure often called the peak signal-to-noise ratio
(PSNR), ${\rm PSNR}=20\log_{10}\left(\nicefrac{\max\left(\hat{\mathbf{x}}\right)}{{\rm {\rm std}\left(\mathbf{x}-\hat{\mathbf{x}}\right)}}\right)$
where ${\rm std}\left(\cdot\right)$ denotes the standard deviation
and; (3) The Structural Similarity Index (SSIM) by \citet{[WanBSS04]}.
The PSNR in decibel (dB), represents a measure that not only contains
RMSE information but also encodes the recovered range. The latter
measure varies between -1 and 1 and the upper bound refers to the
case where the estimated and reference fields are perfectly matched.
The SSIM metric is popular in the image processing community as it
takes into account not only the marginal statistics such as the RMSE
but also the correlation structure between the estimated and reference
(true) image. This metric seems very promising for analyzing the forecast
mismatch with observations in hydro-meteorological studies, especially
when the systematic errors in the large scale features of the predicted
state (e.g., displacement error) might be more dominant than the random
errors; see \citet{[EbtF12a]} for applications of SSIM in rainfall
downscaling.

On average, it was seen that one third of the lost relative energy
of the rainfall reflectivity fields can be restored via the regularized
DS (Table \ref{Tab_1}). The $l_{2}$-regularization led to smoother
results and as the scaling ratio grows, this regularization was almost
incapable to recover the peaks and the correct variability range of
the rainfall field (Figure \ref{Fig_7}). Typically, as expected,
the Huber regularization results were slightly better than the Tikhonov
ones. For large scaling ratios (i.e., > $4\times4$ km) the results
of those methods tended to coincide in terms of the global quality
metrics such the RMSE. However, using the Huber prior, the recovered
range was markedly better than that by the Tikhonov regularization
as reflected in the PSNR metric.

\begin{table}
\noindent \begin{centering}
\begin{tabular}{|c|c|c|c|c|c|c|}
\hline
Metric$^{\dagger}$ & \multicolumn{2}{c|}{Observations} & \multicolumn{2}{c|}{Tikhonov} & \multicolumn{2}{c|}{Huber}\tabularnewline
\hline
\hline
 & $4\times4$ km & $8\times8$ km & $4\times4$ km & $8\times8$ km & $4\times4$ km & $8\times8$ km\tabularnewline
\hline
${\rm RMSE}_{{\rm r}}$ & 0.19 & 0.29 & 0.15 & 0.20 & 0.14 & 0.19\tabularnewline
\hline
${\rm MAE}_{{\rm r}}$ & 0.15 & 0.25 & 0.13 & 0.18 & 0.11 & 0.17\tabularnewline
\hline
SSIM & 0.71 & 0.56 & 0.78 & 0.66 & 0.80 & 0.66\tabularnewline
\hline
PSNR & 23.8 & 19.6 & 26.5 & 23.1 & 27.0 & 24.0\tabularnewline
\hline
\end{tabular}
\par\end{centering}

\caption{Results showing the effectiveness of the proposed regularized DS by
reducing the estimation error and increasing the accuracy of the rainfall
fields. The first two columns refer to the values of the quality metrics
obtained by comparing the constructed low-resolution observations
with true $1\times1$ km reflectivity field. The performance of the
Huber prior is slightly better than the $l_{2}$-regularization, especially
for the small scaling ratios (i.e., $\leq4\times4$ km).$^{\dagger}$
${\rm RMSE}_{r}$: relative root mean squared error; ${\rm MAE}_{r}$
: relative maximum absolute error; SSIM: structural similarity; and
PSNR: peak signal to noise ratio.\textbf{ }\label{Tab_1}}
\end{table}

\section{Regularized Data Fusion }

\subsection{Problem Formulation}

Analogous to the DS problem in the previous section, here we focus
on the formulation of the DF problem. In the DF class of problems,
typically, an improved estimate of the true state is sought from a
series of low-resolution and noisy observations. Let $\mathbf{x}\in\mathbb{R}^{m}$
be the true state of interest while a set of $N$ downgraded measurements
$\mathbf{y}^{i}\in\mathbb{R}^{n_{i}}$, $i=1,\ldots,\, N$, are available
through the following linear observation model
\begin{equation}
\mathbf{y}^{i}=\mathbf{H}^{i}\mathbf{x}+\mathbf{v},
\end{equation}
where $n_{i}\ll m$, $\mathbf{H}^{i}\in\mathbb{R}^{n_{i}\times m}$
and $\mathbf{v}^{i}\sim\mathcal{N}\left(0,\,\mathbf{R}^{i}\right)$
denotes uncorrelated Gaussian error in $\mathbb{R}^{n_{i}}$, $\underset{i\neq j}{\mathbb{E}}\left[\mathbf{v}^{i}\left(\mathbf{v}^{j}\right)^{T}\right]=0$.
Compared to the DS family of problems, DF is more constrained in the
sense that usually there are more equations than the number of unknowns,
$\Sigma_{i}^{N}n_{i}\gg m$, giving rise to an overdetermined linear
system. As previously explained, naturally the linear weighted least
squares (WLS) estimate of the true state, given the series of $N$
observations, amounts to solving the following optimization problem:
\begin{equation}
\hat{\mathbf{x}}=\underset{\mathbf{x}}{{\rm argmin}}\left\{ \frac{1}{2}\sum_{i=1}^{N}\left(\left\Vert \mathbf{y}^{i}-\mathbf{H}^{i}\mathbf{x}\right\Vert _{\left(\mathbf{R}^{i}\right)^{-1}}^{2}\right)\right\} .\label{eq:16}
\end{equation}
Notice that the solution of the above problem not only contains information
about all of the available observations (Fusion) but also, with proper
design of the observation operators, allows us to obtain an estimate
with higher resolution than any of the available observations (Downscaling).
Clearly, the inverse of each covariance matrix in (\ref{eq:16}) plays
the role of the relative contribution or weight of each $\mathbf{y}^{i}$
in the overall cost. In other words, if the elements of covariance
matrix of a particular observation vector are large compared to those
of the other observation vectors, naturally, the contribution of that
observation to the obtained solution would be less significant.

For notational convenience, the above system of equations can be augmented
as follows:
\begin{eqnarray}
\begin{bmatrix}\mathbf{y}^{1}\\
\vdots\\
\mathbf{y}^{N}
\end{bmatrix} & = & \begin{bmatrix}\mathbf{H}^{1}\\
\vdots\\
\mathbf{H}^{N}
\end{bmatrix}\mathbf{x}+\begin{bmatrix}\mathbf{v}^{1}\\
\vdots\\
\mathbf{v}^{N}
\end{bmatrix}\nonumber \\
\Rightarrow\underline{\mathbf{y}} & = & \underline{\mathbf{H}}\mathbf{x}+\underline{\mathbf{v}}\label{eq:17}
\end{eqnarray}
where, the concatenated error vector $\underline{\mathbf{v}}$ has
the following block diagonal covariance matrix,
\begin{equation}
\mathbf{\underline{R}}=\mathbb{E}\left[\mathbf{\underline{\mathbf{v}}}\mathbf{\underline{\mathbf{v}}}^{T}\right]=\begin{bmatrix}\mathbf{R}^{1} &  & 0\\
 & \ddots\\
0 &  & \mathbf{R}^{N}
\end{bmatrix}.
\end{equation}
Therefore, the DF problem can be recast as the classic problem of
estimating the true state from the augmented observation model of
$\underline{\mathbf{y}}=\underline{\mathbf{H}}\mathbf{x}+\underline{\mathbf{v}}$.
Setting the gradient of the cost function in equation (\ref{eq:16})
to zero, yields the following linear system:
\begin{equation}
\left(\underline{\mathbf{H}}^{T}\mathbf{\underline{R}}^{-1}\mathbf{\underline{H}}\right)\hat{\mathbf{x}}=\mathbf{\underline{H}}^{T}\mathbf{\underline{R}}^{-1}\mathbf{\underline{y}}.
\end{equation}
In this case, the left hand side $\underline{\mathbf{H}}^{T}\mathbf{\underline{R}}^{-1}\mathbf{\underline{H}}$
is likely to be very ill-conditioned giving rise to an \emph{unstable}
LS solution, highly sensitive to any perturbation in the observation
vector in the right-hand side (Figure \ref{Fig_8}c) \citep[see, e.g.,][]{[ElaF97],[Han10]}.

One possible remedy for stabilizing the solution is regularization.
Recalling the formulation discussed in the previous section, a general
regularized form of the DF problem can be written as
\begin{equation}
\hat{\mathbf{x}}=\underset{\mathbf{x}}{{\rm argmin}}\left\{ \frac{1}{2}\left\Vert \mathbf{\underline{y}}-\underline{\mathbf{H}}\mathbf{x}\right\Vert _{\mathbf{\underline{R}}^{-1}}^{2}+\lambda\,\psi_{\mathbf{L}}\left(\mathbf{x}\right)\right\} ,\label{eq:20}
\end{equation}
where the convex function $\psi_{\mathbf{L}}\left(\mathbf{x}\right)$
can take different penalty norms such as: the smooth Tikhonov $\left\Vert \mathbf{L}\mathbf{x}\right\Vert _{2}^{2}$;
the non-smooth $l_{1}$-norm $\left\Vert \mathbf{L}\mathbf{x}\right\Vert _{1}$;
and the smooth Huber-norm $\left\Vert \mathbf{L}\mathbf{x}\right\Vert _{{\rm Hub}}$.

For the case of the linear Tikhonov regularization, computation of
the solution requires to invert the Hessian ($\mathbf{\underline{H}}^{T}\mathbf{\underline{R}}^{-1}\mathbf{\underline{H}}+2\lambda\,\mathbf{L}^{T}\mathbf{L}$)
of the objective function similar to equation (\ref{eq:6}). Analogous
to the DS problem, the covariance of the posterior distribution is
$\left(\mathbf{\underline{H}}^{T}\mathbf{\underline{R}}^{-1}\mathbf{\underline{H}}+2\lambda\,\mathbf{L}^{T}\mathbf{L}\right)^{-1}$.

Based on the selected type of regularization, statistical interpretation
of the DF regularized class of problems is also similar to what was
explained in Section 3.2. In other words, given the augmented classic
observation model in (\ref{eq:17}), it is easy to see that the solution
of (\ref{eq:16}) is the ML estimator while (\ref{eq:20}) can be
interpreted as the MAP estimator with a prior density depending on
the form of the regularization term.

\subsection{Example on Rainfall DF}

\begin{figure}[t]
\noindent \begin{centering}
\includegraphics[scale=0.7]{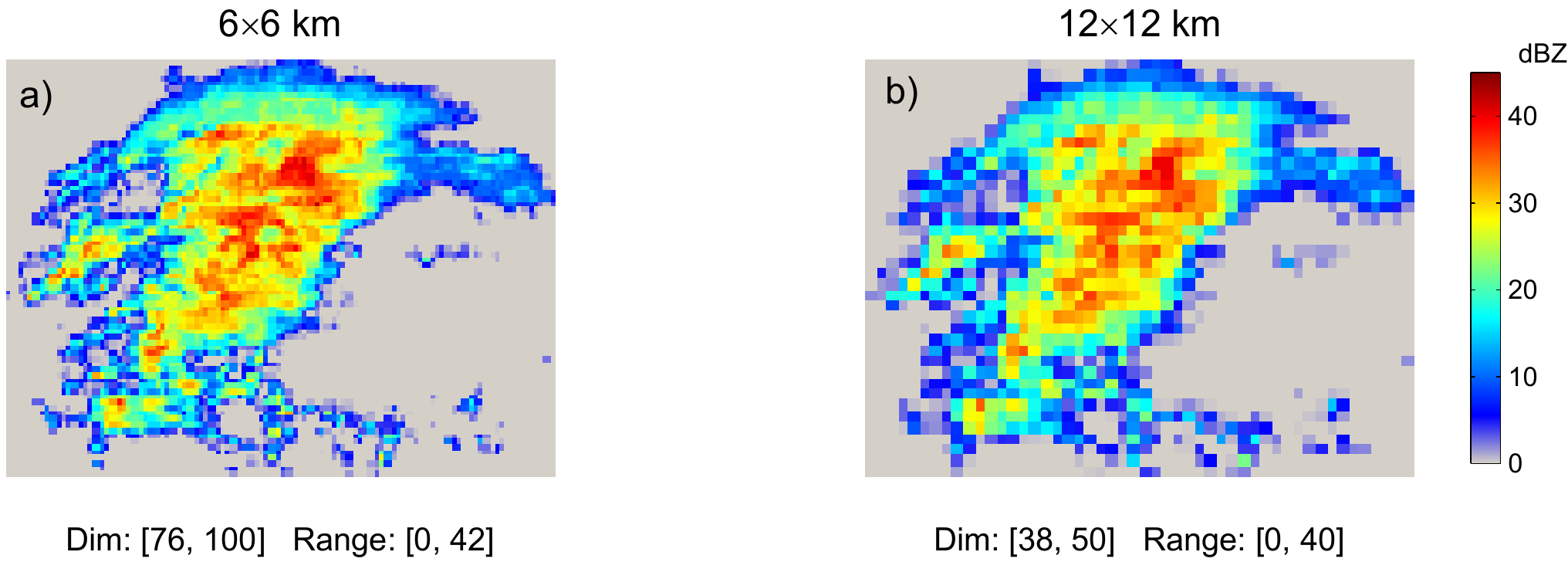}
\par\end{centering}

\noindent \begin{centering}
\includegraphics[scale=0.7]{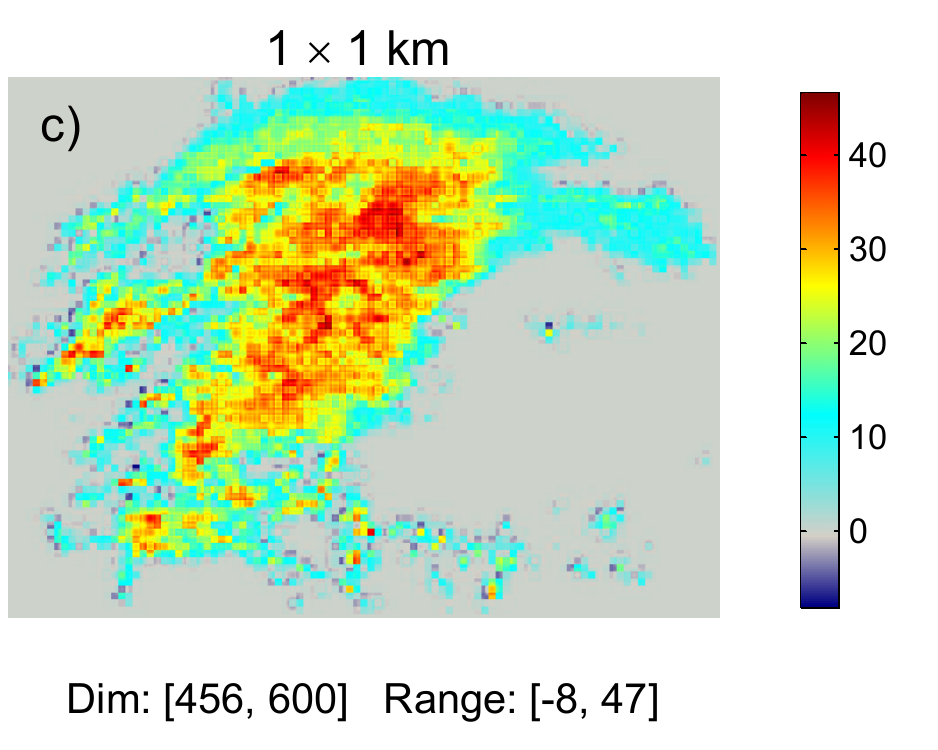} ~~~~~~~\includegraphics[scale=0.7]{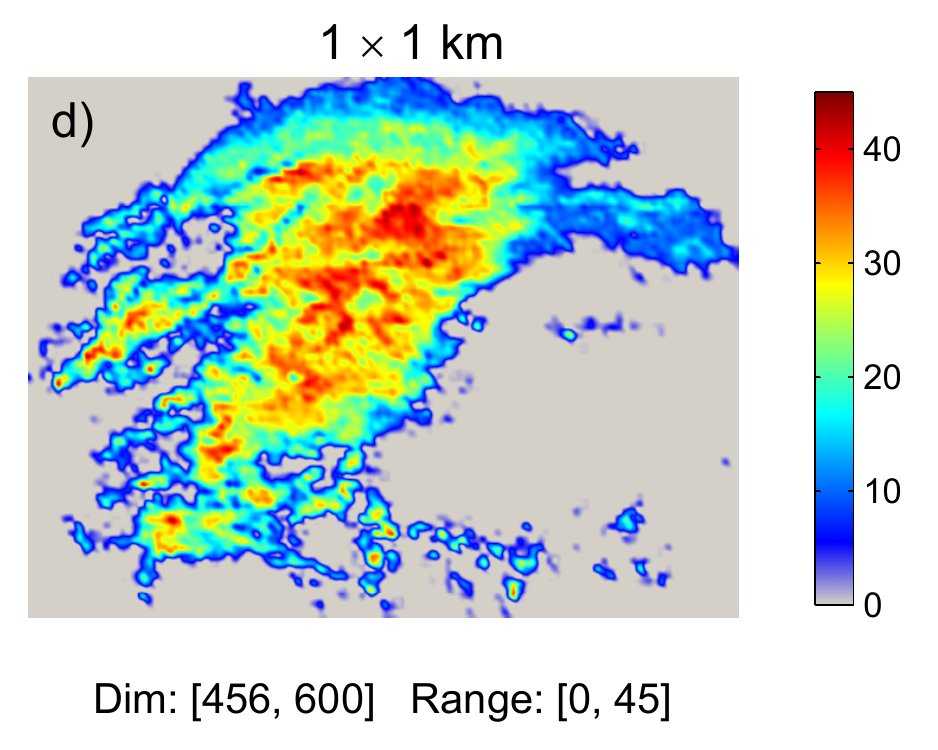}
\par\end{centering}

\caption{Data Fusion and Downscaling of multi-sensor remotely sensed rainfall
reflectivity fields using the Huber regularization. (a-b) Reconstructed
low-resolution and noisy rainfall observations at scale $6$ and $12$
km in grid spacing. (c) The results of the WLS solution in (\ref{eq:16}),
and (d) the solution of Huber regularized DF with $\lambda=1{\rm e}$-3
and $T=1{\rm e}$-2. \label{Fig_8}}
\end{figure}

To quantitatively analyze the effectiveness of the regularized DF
for rainfall data, we reconstructed two synthetic low-resolution and
noisy observations from the original high-resolution NEXRAD reflectivity
snapshot. To resemble different sensor constraints we chose different
smoothing and downsampling operations for each of the reconstructed
field. The first observation field $\mathbf{y}_{1}$ was produced,
at resolution $6\times6$ km, using a simple averaging filter of size
$6\times6$ followed by a downsampling ratio of $s_{c}=6$. Analogously,
the second field $\mathbf{y}_{2}$ was generated at scale $12\times12$
km using a Gaussian smoothing kernel of size the $12\times12$ with
a standard deviation of $4$. A white Gaussian noise, with standard
deviation of $1{\rm e}$-2 and $2{\rm e}$-2 was also added, respectively,
to resemble the measurement random error. Roughly speaking, this selection
of the error magnitudes implies that the degree of confidence (relative
weight) on the observation at $6\times6$ km is twice as large as
for the one at $12\times12$ km scale. According to the selected smoothing
and downsampling operations, the downgrading operators $\mathbf{H}^{1}$
and $\mathbf{H}^{2}$ are designed to produce a high-resolution field
at the scale of $1\times1$ km. To solve the DF problem, we have used
the same settings for the Gradient Projection (GP) method as explained
in Appendix A.

The solution of the ill-conditioned WLS formulation or the ML estimator
in (\ref{eq:16}) is blocky, out of range and severely affected by
the amplified inverted noise (Figure \ref{Fig_8}c). On the other
hand, the Huber regularization can properly restore a fine-scale and
coherent estimate of the rainfall field. The results show that almost
30\% of the uncaptured subgrid energy of the rainfall field can be
restored through solving the regularized DF problem (Table \ref{Tab_2}).
Improvements of the selected fidelity measures in the DF problem is
more pronounced than the results of the DS experiment. This naturally
arises, because more observations are available in the DF problem
than the DS one and thus, the results are likely to be more accurate.

\begin{table}
\noindent \begin{centering}
\begin{tabular}{|c|c|c|c|}
\hline
Metric & \multicolumn{2}{c|}{Observations} & DF results\tabularnewline
\hline
\hline
 & $6\times6$ km & $12\times12$ km & $1\times1$ km\tabularnewline
\hline
${\rm RMSE}_{{\rm r}}$ & 0.25 & 0.35 & 0.17\tabularnewline
\hline
${\rm MAE}_{{\rm r}}$ & 0.21 & 0.32 & 0.15\tabularnewline
\hline
SSIM & 0.60 & 0.50 & 0.72\tabularnewline
\hline
PSNR & 21.3 & 18.1 & 25.0\tabularnewline
\hline
\end{tabular}
\par\end{centering}

\caption{Values of the selected fidelity metrics in the rainfall DF experiment
using the Huber regularization, see the text for the definitions.
Here, metrics refer to comparison of the low-resolution ($6\times6$
and $12\times12$ km) observations the DF results with the true field
($1\times1$ km).\label{Tab_2}}
\end{table}

\section{Regularized Variational Data Assimilation }

\subsection{Problem Formulation}

Compared to the previously explained problems of downscaling and data
fusion, the data assimilation (DA) problem is more involved in the
sense that we need to address the evolution of a dynamical system
and the available (low-resolution) observations at the same time in
the estimation process. Despite the increased complexity, DA shares
the same principles with the explained formulations of the DS and
DF problems, from the estimation point of view. Here, we briefly explain
the linear 3D and 4D-VAR data assimilation schemes and extend their
formulations to a regularized format. Sample results of the regularized
variational data assimilation problem are illustrated on the estimation
of the initial conditions of the linear heat equation in a 3D-VAR
setting.

The 3D-VAR is a memoryless assimilation method. In other words, at
each time step, the best estimate of the true state or say \emph{analysis}
is obtained based only on the present-time noisy \emph{observations}
and \emph{background} information of the dynamical system. The analysis
is then being used for forecasting the state at the next time step
an so on. Suppose that the true state of interest at discrete time
$t_{k}$ is denoted by $\mathbf{x}_{k}\in\mathbb{R}^{m}$, a single
noisy observation is $\mathbf{y}_{k}\in\mathbb{R}^{n}$, and $\mathbf{x}_{k}^{b}\in\mathbb{R}^{m}$
represents the background state. In the linear 3D-VAR data assimilation
problem, obtaining the \emph{analysis} state $\mathbf{x}_{k}^{a}\in\mathbb{R}^{m}$
amounts to finding the minimum point of the following cost function:

\begin{equation}
\mathcal{J}_{3D}(\mathbf{x}_{k})=\frac{1}{2}\left\Vert \mathbf{x}_{k}^{b}-\mathbf{x}_{k}\right\Vert _{\mathbf{B}^{-1}}^{2}+\frac{1}{2}\left\Vert \mathbf{y}_{k}-\mathbf{H}\mathbf{x}_{k}\right\Vert _{\mathbf{R}^{-1}}^{2}.\label{eq:21}
\end{equation}
In equation (\ref{eq:21}), $\mathbf{B}\in\mathbb{R}^{m\times m}$
is the background error covariance matrix, $\mathbf{H}$ denotes the
observation operator, while the analysis is the optimal solution:
$\mathbf{x}_{k}^{a}={\rm \underset{\mathbf{x}_{k}}{argmin}}\left\{ \mathcal{J}_{3D}(\mathbf{x}_{k})\right\} $.
Here we assume that the observation matrix is translation invariant.
Clearly, this 3D-VAR problem is the WLS problem which has the following
analytic solution
\begin{equation}
\mathbf{x}_{k}^{a}=\left(\mathbf{B}^{-1}+\mathbf{H}^{T}\mathbf{R}^{-1}\mathbf{H}\right)^{-1}\left(\mathbf{B}^{-1}\mathbf{x}_{k}^{b}+\mathbf{H}^{T}\mathbf{R}^{-1}\mathbf{y}_{k}\right).
\end{equation}
Because the error covariance matrices are positive definite, the matrix
$\mathbf{B}^{-1}+\mathbf{H}^{T}\mathbf{R}^{-1}\mathbf{H}$ is always
positive definite and hence invertible. Thus, solution of the 3D-VAR
requires no rank or dimension assumption on $\mathbf{H}$. However,
this problem might be very ill-conditioned depending on the architecture
of the covariance matrices and the measurement operator.

The classic 4D-VAR is, indeed, an extension to the explained 3D-VAR
which exploits the temporal memory of the system by constraining the
solution (analysis) to the underlying dynamics. Assuming that we are
interested in estimating the initial condition of a dynamical model
at previous discrete time $t_{k}$, to be used for an improved forecast
in future time. The 4D-VAR assimilation method is formulated in such
a way that uses the initial background state $\mathbf{x}_{k}^{b}$
and also the observations within a finite discrete time interval $[t_{k},\ldots,\, t_{k+T}]$.
Accordingly, this assimilation method amounts to obtaining the minimum
point of the following cost function:
\begin{equation}
\mathcal{J}_{4D}(\mathbf{x}_{k})=\frac{1}{2}\left\Vert \mathbf{x}_{k}^{b}-\mathbf{x}_{k}\right\Vert _{\mathbf{B}^{-1}}^{2}+\frac{1}{2}\sum_{i=k}^{k+T}\left(\left\Vert \mathbf{y}_{i}-\mathbf{H}\mathbf{x}_{i}\right\Vert _{\mathbf{R}_{i}^{-1}}^{2}\right),
\end{equation}
while constraining the solution to the underlying dynamics by assuming
that $\mathbf{x}_{i}=\mathbf{M}_{k,i}\,\mathbf{x}_{k}$, where $t_{i}\geq t_{k}$.
The model operator $\mathbf{M}_{k,i}$ is a discrete linear representation
of the system dynamics that evolves the state from $t_{k}$ to $t_{i}$
\citep[e.g.,][]{[CouT90],[Dal93],[Zup93],[Kal03]}. Note that, in
the above formulation it is implicitly assumed that the background
and all measurement errors are mutually uncorrelated. Clearly, the
linear 4D-VAR optimization contains a background cost which measures
the weighted Euclidean distance of the true initial condition to the
background state at the beginning of the interval and an accumulated
cost, associated with all of the noisy observations within the selected
time interval. Taking into account the imposed constraint by the model
operator, the 4D-VAR cost function can be recast as,
\begin{equation}
\mathcal{J}_{4D}(\mathbf{x}_{k})=\frac{1}{2}\left\Vert \mathbf{x}_{k}^{b}-\mathbf{x}_{k}\right\Vert _{\mathbf{B}^{-1}}^{2}+\frac{1}{2}\sum_{i=k}^{K+T}\left(\left\Vert \mathbf{y}_{i}-\mathbf{H}\mathbf{M}_{k,\, i}\,\mathbf{x}_{k}\right\Vert _{\mathbf{R}_{i}^{-1}}^{2}\right),\label{eq:24}
\end{equation}
where its optimal solution, $\mathbf{x}_{k}^{a}={\rm \underset{\mathbf{x}_{k}}{argmin}}\left\{ \mathcal{J}_{4D}(\mathbf{x}_{k})\right\} $,
is the analysis state at $k^{th}$ time-step. Thus, it is easy to
see that, the linear 3D and 4D-VAR problems are in the category of
the classic WLS problems which might be very ill-conditioned.

Analogous to the previous discussions, the generic regularized form
of the linear 3D-VAR under the predetermined $\mathbf{L}$-transformation
might be considered as follows:
\begin{equation}
\mathbf{x}_{k}^{a}=\underset{\mathbf{x}_{k}}{{\rm argmin}}\left\{ \mathcal{J}_{3D}(\mathbf{x}_{k})+\lambda\,\psi_{\mathbf{L}}\left(\mathbf{x}_{k}\right)\right\} ,\label{eq:25}
\end{equation}
where $\psi_{\mathbf{L}}\left(\mathbf{x}\right)$ can take any of
the explained regularization penalty functions including: the smooth
Tikhonov $\left\Vert \mathbf{L}\mathbf{x}\right\Vert _{2}^{2}$; the
non-smooth $l_{1}$-norm $\left\Vert \mathbf{L}\mathbf{x}\right\Vert _{1}$;
and the smooth Huber-norm $\left\Vert \mathbf{L}\mathbf{x}\right\Vert _{{\rm Hub}}$.
It is easy to realize that the regularized 4D-VAR estimation of the
the initial condition can also take the following form:
\begin{equation}
\mathbf{x}_{k}^{a}=\underset{\mathbf{x}_{k}}{{\rm argmin}}\left\{ \mathcal{J}_{4D}(\mathbf{x}_{k})+\lambda\,\psi_{\mathbf{L}}\left(\mathbf{x}_{k}\right)\right\} .
\end{equation}
In the above regularized formulations, the solution not only becomes
close to the background and observations, in the weighted Euclidean
sense, but it is also enforced to follow a regularity imposed by the
$\psi_{\mathbf{L}}\left(\mathbf{x}\right)$. Here, we emphasize that
the regularization typically yields a stable and improved solution
with less uncertainty; however, this gain comes at the price of introducing
bias in the solution whose magnitude can be kept small, by proper
selection of the regularizer \citep{[Han10]}.

\subsection{Statistical Interpretation}

Statistical interpretation of the classic variational DA problems
is a bit tricky compared to the DS and DF class of problems, mainly
because of the involvement of the background information in the cost
function. \citet{[Lor86]} derived the 3D-VAR cost function using
Bayes theorem and called it the ML estimator \citep[see, e.g.,][]{[Lor88],[BouC02]}.
More recently, it has been argued that the 4D-VAR, and thus as a special
case the 3D-VAR cost function, can be interpreted via the Bayesian
MAP estimator \citep{[JohNH05],[FreNB10],[Nic10]}. For notational
convenience, here we only explain the statistical interpretation of
the 3D-VAR and its regularized version which can be easily generalized
for the case of the 4D-VAR problem.

As discussed earlier, the ML estimator is basically a frequentist
view to estimate the most likely value of an unknown deterministic
variable from an (indirect) observation with random nature. The ML
estimator intuitively requires to find the state that maximizes the
likelihood function as $\hat{\mathbf{x}}_{ML}={\rm \underset{\mathbf{x}}{argmax}}\, p\left(\mathbf{y}|\mathbf{x}\right)$.
Let us assume that, at time step $t_{k}$, the background $\mathbf{x}_{k}^{b}$
is just a (random) realization of the true deterministic state $\mathbf{x}_{k}$.
In other words, we consider $\mathbf{x}_{k}^{b}=\mathbf{x}_{k}+\mathbf{w}$,
where the error $\mathbf{w}$ can be well explained by a zero mean
Gaussian density $\mathcal{N}\left(0,\,\mathbf{B}\right)$, uncorrelated
with the observation error, $\mathbb{E}\left[\mathbf{w}\mathbf{v}^{T}\right]=0$.
Here, the background state is treated similar to an observation with
random nature. Thus, let us recast the problem of obtaining the analysis
as a classic linear inverse problem by augmenting the available information
in the from of $\underline{\mathbf{y}}=\underline{\mathbf{H}}\mathbf{x}_{k}+\mathbf{\underline{v}}$,
where $\underline{\mathbf{y}}=\left[\left(\mathbf{x}_{k}^{b}\right)^{T},\,\mathbf{y}_{k}^{T}\right]^{T}$
, $\underline{\mathbf{H}}=\left[\mathbf{I},\,\mathbf{H}^{T}\right]^{T}$,
and $\underline{\mathbf{v}}\sim\mathcal{N}\left(0,\,\mathbf{\underline{R}}\right)$
with the following block diagonal covariance matrix

\begin{equation}
\underline{\mathbf{R}}=\begin{bmatrix}\mathbf{B} & 0\\
0 & \mathbf{R}
\end{bmatrix}.
\end{equation}

Notice that $\mathbf{\underline{R}}$ is block diagonal because the
background and observation errors are uncorrelated. Following the
augmented representation and applying $-\log\left(\cdot\right)$,
we have $-\log\, p(\mathbf{\underline{y}}|\mathbf{x}_{k})\propto\nicefrac{1}{2}(\mathbf{\underline{y}}-\mathbf{\underline{H}}\mathbf{x}_{k})^{T}\mathbf{\underline{R}}^{-1}(\underline{\mathbf{y}}-\mathbf{\underline{H}x}_{k})$
and thus it is easy to see that the ML estimator in terms of the augmented
observations, $\mathbf{x}_{k}^{a}={\rm \underset{\mathbf{x}_{k}}{argmax}}\, p\left(\mathbf{\underline{y}}|\mathbf{x}_{k}\right)$,
is equivalent to minimizing the 3D-VAR cost function in (\ref{eq:21}).
Therefore, following this statistical interpretation, the classic
3D-VAR, can be derived via the frequentist ML estimator.

On the other hand, from the Bayesian perspective, the state of interest
and the available observations are considered to be random and the
MAP estimator is the optimal point which maximizes the posterior density
as $\hat{\mathbf{x}}_{MAP}={\rm \underset{\mathbf{x}}{argmax}}\, p\left(\mathbf{x}|\mathbf{y}\right)$.
Let us assume a priori that the (random) state of interest has a Gaussian
density with the mean $\mathbf{x}_{b}$ and covariance $\mathbf{B}$,
that is $p(\mathbf{x}_{k})\sim\mathcal{N}(\mathbf{x}_{k}^{b},\,\mathbf{B})$.
More formally, this assumption implies that the deterministic background
is the central (mean) forecast and is related to the random true state
via $\mathbf{x}_{k}=\mathbf{x}_{k}^{b}+\mathbf{w}$, where $\mathbf{w}\sim\mathcal{N}\left(0,\,\mathbf{B}\right)$.
Therefore, using Bayes theorem; see equation (\ref{eq:12}), it immediately
follows that the 3D-VAR is the MAP estimator with the assumed Gaussian
prior for the true state, $\mathbf{x}_{k}^{a}={\rm \underset{\mathbf{x}_{k}}{argmax}}\, p\left(\mathbf{x}_{k}|\mathbf{y}\right)$.

In conclusion, the regularized 3D-VAR in (\ref{eq:25}) might be interpreted
as the MAP estimator, $\mathbf{x}_{k}^{a}={\rm \underset{\mathbf{x}_{k}}{argmax}}\, p\left(\mathbf{x}_{k}|\mathbf{\underline{y}}\right)$,
with the prior density, $p(\mathbf{x}_{k})\propto\lambda\,\psi_{\mathbf{L}}\left(\mathbf{x}_{k}\right)$,
when we follow the frequentist approach and use the augmented notation
to interpret the classic 3D-VAR as the ML estimator. On the other
hand, taking the MAP interpretation for the classic 3D-VAR, the regularized
version might be understood as the MAP estimator which also accounts
for an extra and independent prior on the distribution of the state
under the $\mathbf{L}$ transformation.

\subsection{Heat Equation Example}

The promise of the proposed regularized 3D-VAR data assimilation methodology,
is shown via assimilating noisy observations into the dynamics of
heat equation with top-hat initial condition. Specifically, we constructed
noisy background and noisy-low-resolution observations of the top-hat
initial condition and then demonstrated the effectiveness of a proper
regularization on the quality of the obtained analysis and forecast
states. In the assimilation cycle, we obtained the analyses using
the classic and regularized 3D-VAR assimilation methods and then,
we examined those analysis states to obtain the forecast state at
the next time step. Then the computed analysis and forecast states
are compared with their available ground-truth counterparts. Although
the heat equation has a diffusive nature and is not sensitive to its
initial condition, the provided examples are very illustrative about
the role of regularization on the quality of solutions.

For a space-time representation of a 1D scalar quantity $x\left(s,\, t\right)$,
the well-known heat equation is
\begin{eqnarray}
\frac{\partial x(s,\, t)}{\partial t} & = & \gamma\nabla^{2}x(s,\, t)\\
x(s,\,0) & = & x_{0}(s),\nonumber
\end{eqnarray}
where $\infty<s<\infty,\,\,0<t<\infty$. For mathematical treatment,
let us assume that $\gamma=1\,\left[\unitfrac{{\rm L}^{2}}{T}\right]$.
It is well understood that the general solution of the heat equation
at time $t$ is then given by the convolution of the initial condition
with the fundamental solution (kernel) as
\begin{equation}
x(s,\, t)=\int K\left(s-r,\, t\right)\, x_{0}(r)\, dr,
\end{equation}
where
\begin{equation}
K(s,\, t)=(4\pi t)^{-m/2}\exp\left(\frac{-\left|s\right|^{2}}{4t}\right).
\end{equation}
We can see that $x\left(s,\, t\right)$ is obtained via convolution
of the initial condition by a Gaussian kernel with the standard deviation
of $\sigma=\sqrt{2t}$. Clearly, estimation of the initial condition
$x_{0}(s)$ only from the diffused and possibly noisy observations
$x(s,\, t)$ is an ill-posed deconvolution problem (see equation \ref{eq:1}).

To reconstruct a 3D-VAR assimilation experiment, we assume that the
true initial condition in discrete space is a vector with 256-elements
($\mathbf{x}\in\mathbb{R}^{m}$ where $m=256$) as follows:
\begin{equation}
\mathbf{x}_{0}=\begin{cases}
2 & \,\,112\leq x_{i}\leq144\\
1 & {\rm \,\,\,\,\,\, otherwise},
\end{cases}
\end{equation}
the so-called top-hat initial condition. We added a white Gaussian
noise with $\sigma_{w}=0.05$ to the true initial condition for reconstructing
the background state $\mathbf{x}_{0}^{b}$ for the assimilation experiment.

\begin{figure}[t]
\noindent \begin{centering}
\includegraphics[scale=0.7]{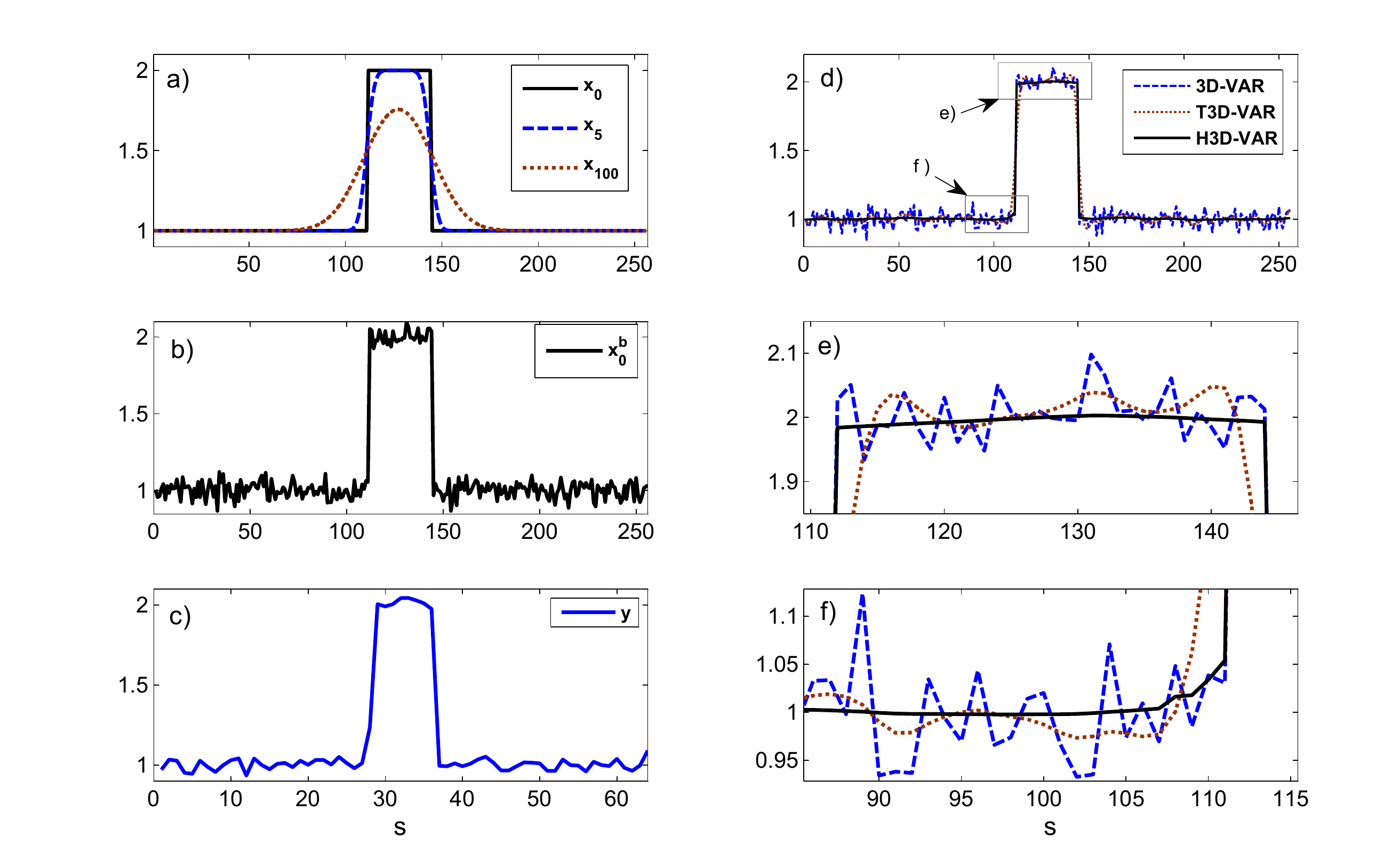}
\par\end{centering}

\caption{(a) The true initial condition $\mathbf{x}_{0}$ and the results of
the heat equation at $t=5$ and 100 $[{\rm T}]$, obtained from convolution
of the initial condition with the fundamental Gaussian kernel. (b)
The reconstructed background state by adding a white noise with $\sigma_{w}=0.05$.
(c) The low-resolution and noisy observation with $\sigma_{v}=0.03$.
(d) The results of the classic 3D-VAR and the regularized version
using the Tikhonov (T3D-VAR) and the Huber (H3D-VAR) regularizations
(see equation \ref{eq:25}). (e-f) Magnified parts of of the graphs
in (d) over the shown zooming windows. \label{Fig_9}}
\end{figure}

We assumed that the observation vector is a downgraded version of
the true state, while the sensor can only capture the mean of every
four neighbor elements of the true state. In other words, the observation
is a noisy and low-resolution version of the true state with one quarter
of its size (Figure \ref{Fig_9}). To this end, using the linear model
in (\ref{eq:2}), we employed the following architecture for the observation
operator:
\begin{equation}
\mathbf{H}=\frac{1}{4}\begin{bmatrix}1\,1\,1\,1 & 0\,0\,0\,0 & \cdots & 0\,0\,0\,0\\
0\,0\,0\,0 & 1\,1\,1\,1 & \cdots & 0\,0\,0\,0\\
\vdots & \vdots & \vdots & \vdots\\
0\,0\,0\,0 & 0\,0\,0\,0 & \cdots & 1\,1\,1\,1
\end{bmatrix}\in\mathbb{R}^{n\times m},
\end{equation}
and an $n$-element white Gaussian observation error with $\sigma_{v}=0.03$,
where $n=64$.

The top-hat initial condition is selected to emphasize the role of
regularization, especially those with linear penalization (i.e., the
Huber or $l_{1}$-norm ). Clearly, the first order derivative of the
above initial condition is very sparse. In other words, the first
order derivative is zero everywhere on its domain except at the location
of the two jumps, resembling a heavy-tailed and sparse statistical
distribution. This underlying structure prompts us to use the $l_{1}$-like
regularization and first order differencing operator for the $\mathbf{L}$-transform
in (\ref{eq:25}), as follows:
\begin{equation}
\mathbf{L}=\begin{bmatrix}\begin{array}{rrrrrr}
-1 & 1 & 0 & \cdots & 0 & 0\\
0 & -1 & 1 & \cdots & 0 & 0\\
\vdots & \vdots & \vdots & \vdots & \vdots & \vdots\\
0 & 0 & 0 & \cdots & -1 & 1
\end{array}\end{bmatrix}\in\mathbb{R}^{(m-1)\times m}.
\end{equation}
Note that, instead of incorporating a measure of curvature in the
regularization term to impose smoothness on the solution, here we
seek a solution with minimize total variation.

Figure \ref{Fig_9} shows the inputs of the assimilation experiment
(left panels) and the results of the analysis cycle (right panels)
using the classic versus the regularized 3D-VAR estimators. In this
example, it is clear that the classic solution overfits, while slightly
damps the noise. Indeed, the 3D-VAR is unable to effectively damp
the high-frequency error components and impose the underlying regularity
of the true state, because, its cost function is only about minimizing
the weighted squared error. On the other hand, in the regularized
assimilation methods not only the error term, but also a cost associated
with the state regularity is also minimized. The Tikhonov regularization
(T3D-VAR), i.e.,$\psi_{\mathbf{L}}\left(\mathbf{x}\right)=\left\Vert \mathbf{Lx}\right\Vert _{2}^{2}$
led to a smoother result compared to the classic one with slightly
better error statistics. However, result of the Huber regularization
(H3D-VAR), i.e., $\psi_{\mathbf{L}}\left(\mathbf{x}\right)=\left\Vert \mathbf{Lx}\right\Vert _{{\rm Hub}}$,
is the best. The rapidly varying noisy components are effectively
damped in this regularization, while the sharp jump discontinuities
have been preserved better than the T3D-VAR. The quantitative metrics
in Table \ref{Tab_3}, indicate that in the analysis cycle, the RMSE
and MAE metrics are improved dramatically, up to 85\% in the H3D-VAR
scheme.

\begin{table}[b]
\noindent \begin{centering}
\begin{tabular}{|c|c|c|c|c|c|c|}
\hline
Cycle & \multicolumn{3}{c|}{RMSE} & \multicolumn{3}{c|}{MAE}\tabularnewline
\hline
\hline
 & 3D-VAR & T3D-VAR & H3D-VAR & 3D-VAR & T3D-VAR & H3D-VAR\tabularnewline
\hline
A & 0.0475 & 0.0397 & 0.0067 & 0.0376 & 0.0317 & 0.0043\tabularnewline
\hline
F & 0.0090 & 0.0088 & 0.0043 & 0.0071 & 0.0070 & 0.0033\tabularnewline
\hline
\end{tabular}
\par\end{centering}

\caption{The root mean squared error (RMSE) and the mean absolute error (MAE)
for the studied classic and regularized 3D-VAR in the analysis cycle
(A) and forecast step (F) . \label{Tab_3} }
\end{table}

\begin{figure}
\noindent \begin{centering}
\includegraphics[scale=0.65]{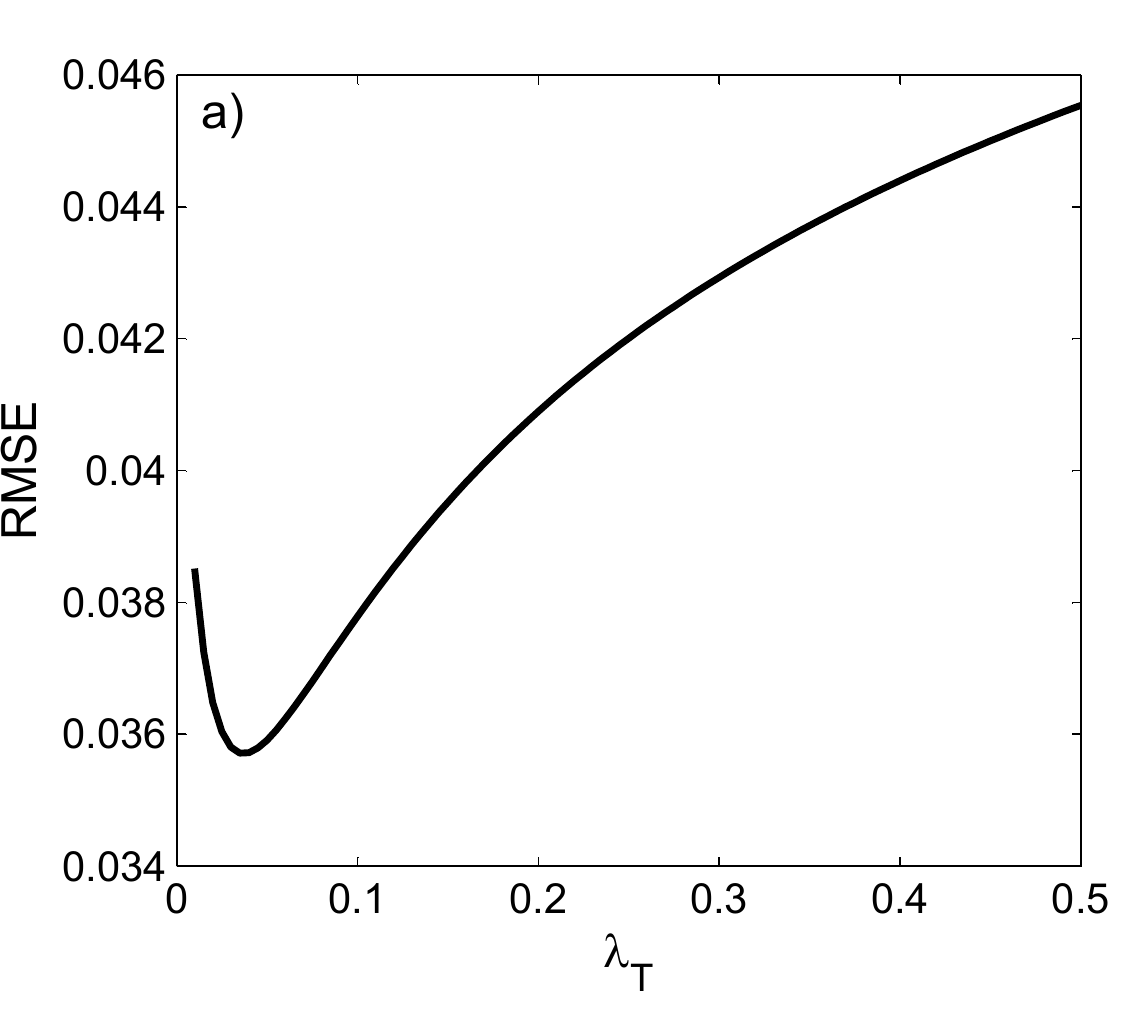}\includegraphics[scale=0.65]{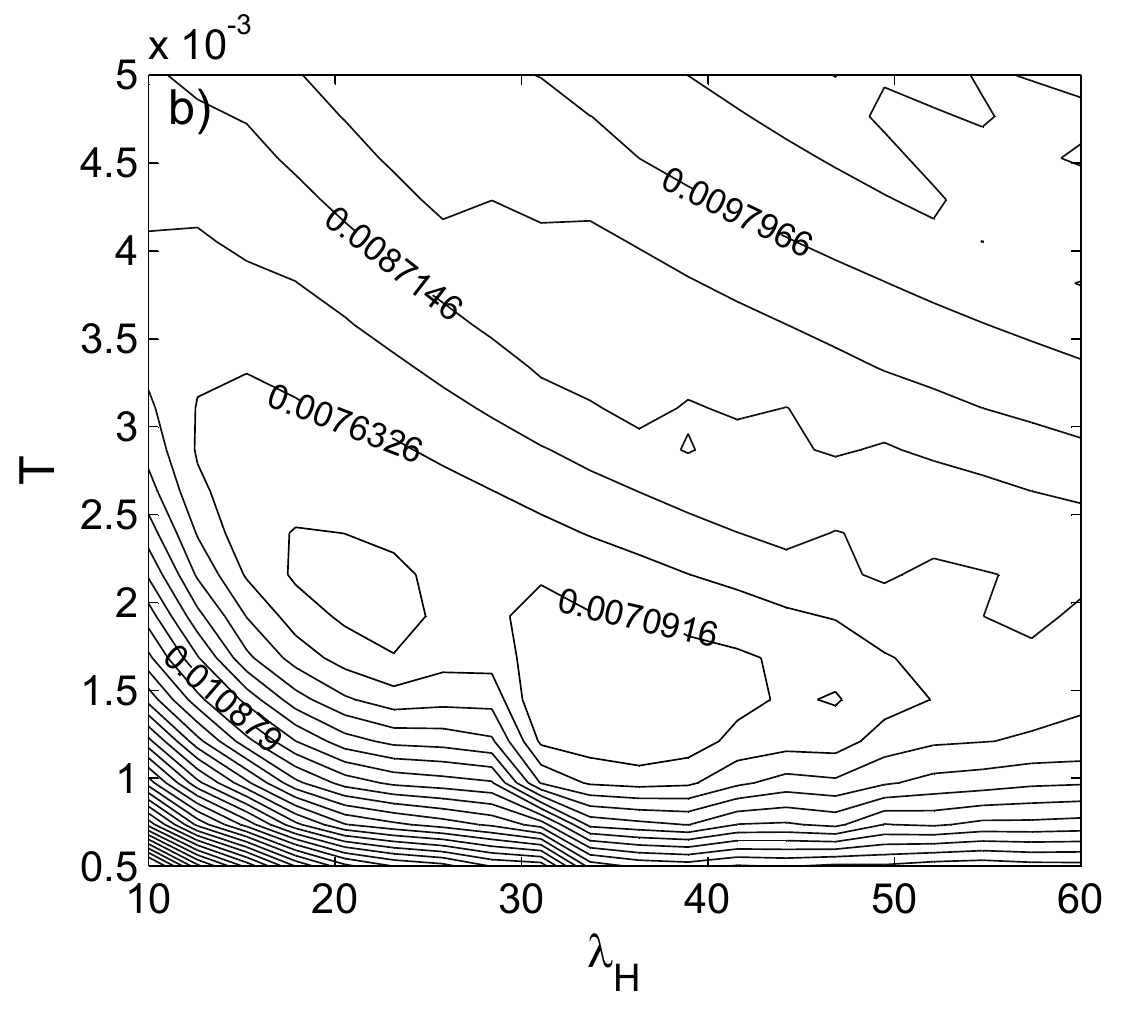}
\par\end{centering}

\caption{(a) Root mean squared error (RMSE) of the implemented T3D-VAR as a
function of the regularizer $\lambda_{T}$. (b) RMSE surface for the
H3D-VAR experiment with different choices of the regularizer$\lambda_{H}$
and the threshold value $T$ of the Huber-norm. Clearly, depending
on the choice of the regularization method, the strength of the regularizer
(magnitude of $\lambda$) might be markedly different. \label{Fig_10}}
\end{figure}

As previously explained, there is no unique and universally accepted
methodology for automated selection of the regularization parameters,
namely $\lambda$ and $T$. Here, to select the best parameters in
the above assimilation examples, we performed a few trial and error
experiments. In other words, over a feasible range of parameter values,
we computed the solutions of the regularized assimilation methods
and obtained the RMSE measure by comparing those solutions with the
true initial condition $\mathbf{x}_{0}$. Figure \ref{Fig_10} shows
the RMSE for different choices of regularization parameters in both
T3D-VAR and H3D-VAR. Note that, the true initial condition is definitely
not available in practice; however, here we picked the optimal values
of the regularization parameters in the RMSE sense for comparison
purposes and demonstrating the importance of a proper regularization.
In the T3D-VAR, as expected, larger values of $\lambda$ typically
damp rapidly varying error components of the noisy background and
observation; however, they may give rise to an overly smooth solution
with larger bias and RMSE (Figure \ref{Fig_10}a). Here, for the T3D-VAR
experiment, we picked the value $\lambda_{T}=0.05$ associated with
the minimum RMSE (Figure \ref{Fig_9}). In the H3D-VAR, in addition
to the regularizer $\lambda_{H}$, we also need to choose the optimal
thresholding value $T$ of the Huber-norm. A contour plot of the RMSE
values for different choices of $\lambda_{H}$ and $T$ is shown in
Figure \ref{Fig_10}b. By inspection, we roughly picked $\lambda_{H}=$35
and $T=1.5{\rm e}$-3 for the H3D-VAR assimilation experiment presented
in Figure \ref{Fig_9}.

\begin{figure}[t]
\noindent \begin{centering}
\includegraphics[scale=0.75]{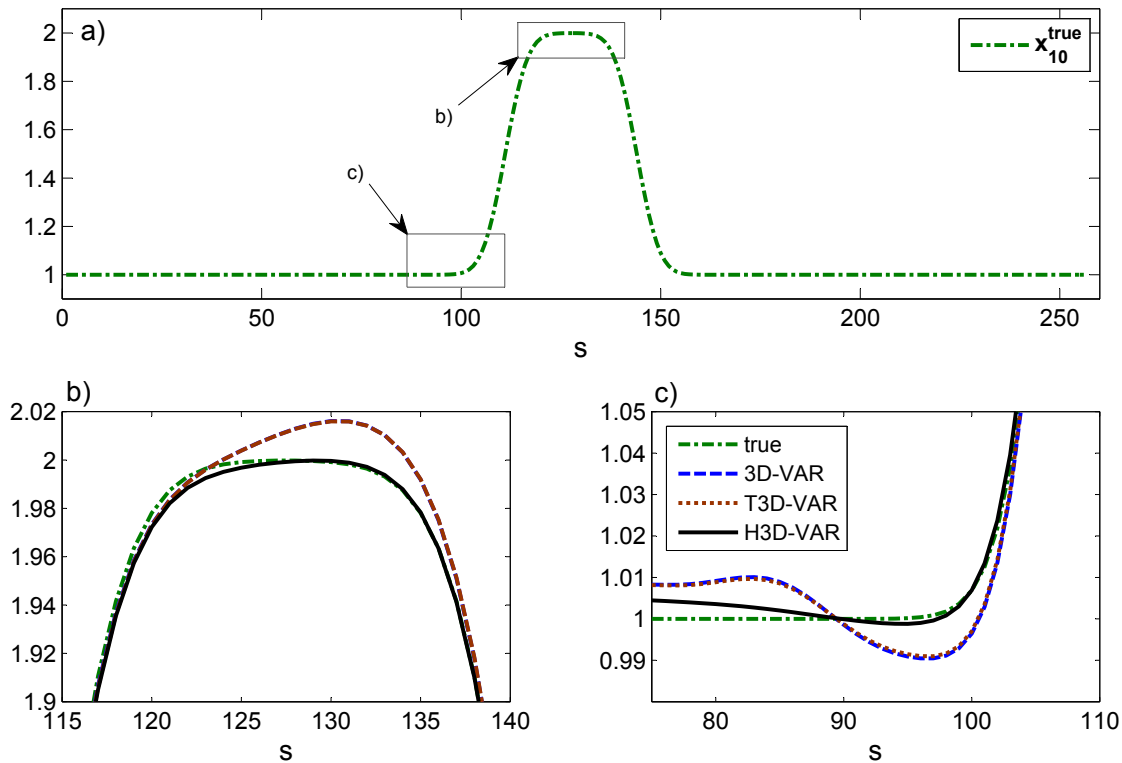}
\par\end{centering}

\caption{(a) True forecast based on temporal evolution of the top-hat initial
condition, Figure \ref{Fig_9}a, under the heat equation at $t=10\,\left[{\rm T}\right]$.
(b-c) Magnified windows showing the forecast quality using different
3D-VAR assimilation methods. Ineffective error removal by the classic
3D and T3D-VAR at the analysis cycle produced large-scale correlated
error in the forecast profile, while this problem is less substantial
in the result of the H3D-VAR (see, Table \ref{Tab_3}). \label{Fig_11}}
\end{figure}

The main purpose of the DA process is, indeed, to increase the quality
of the forecast. Given the initial-time analysis state, we can forecast
the profile of the scalar quantity, $x\left(s,\, t\right)$, at any
future time step through the heat equation. One important property
of the heat equation is its diffusivity. In other words, naturally,
noisy components and rapidly varying perturbations are damped but
become more correlated as the profile of the initial condition evolves.
Thus, rapidly-varying uncorrelated error components become low-varying
and correlated features which their detection and removal are naturally
more difficult than the uncorrelated ones. Figure \ref{Fig_11}a,
shows the forecast profile at $t=10\,\left[{\rm T}\right]$, obtained
by convolution of the initial profile with a Gaussian kernel having
standard deviation of $\sigma=\sqrt{2\times10}$. The results indicate
the importance of using a proper regularization on the quality of
the forecast in the simple heat equation. The forecasts based on the
classic 3D-VAR and the T3D-VAR almost coincide while the regularized
result is marginally better. This behavior arises because neither
of those methods could properly eliminate the noisy features in the
analysis cycle and hence low-varying error components are appeared
in the forecast profile. However, the quality metrics in Table \ref{Tab_3}
indicate that using the H3D-VAR, the RMSE and MAE of the forecast
are almost improved more than 50\% compared to the other methods.

\section{Conclusion }

In this paper, we presented a new direction in approaching hydrometeorological
estimation problems by taking into account important continuity and
statistical properties of the underlying states such as the presence
of sharp jumps, isolated singularities (i.e., local extremes) and
statistical sparsity in the derivative space. We started by explaining
the concept of regularization and discussed about the common points
of the hydrometeorological problems of downscaling (DS), data fusion
(DF), and data assimilation (DA) as discrete linear inverse problems.
We argued about the importance of proper regularization which not
only makes an inverse problem well-posed but also imposes the desired
regularity and statistical property on the solution. Regularization
methods were theoretically linked to the underlying statistical structure
of the states and it was shown how the statistical information about
the density of the state, or its derivative, can be used for proper
selection of the regularization method. Specifically, we emphasized
three types of regularization, namely, the Tikhonov, $l_{1}$-norm
and Huber regularizations. We theoretically showed that these methods
are statistically equivalent to the maximum a posteriori (MAP) estimator
while respectively assuming the Gaussian, Laplace and Gibbs prior
density for the state. It was argued that piece-wise continuity of
the state and the presence of frequent jumps are often translated
into heavy-tailed distributions in the derivative space that favor
the use of $l_{1}$-regularization.

Informed by the non-Gaussian distribution of rainfall derivatives,
the effectiveness of the regularized DS and DF problems was tested
via analysis of remotely sensed precipitation fields. Then the performance
of the regularized DA was studied via assimilating noisy observations
into evolution of the heat equation. We showed that regularized assimilation
methods outperform the classic 3D-VAR method, especially for the case
where the initial condition exhibits a sparse distribution in the
derivative space (e.g., first order derivative of the top-hat initial
condition).

The presented frameworks can be potentially applied to other hydrometeorological
problems, such as soil moisture downscaling and fusion. Clearly, proper
selection of the regularization method requires careful statistical
analysis of the underlying state of interest. Moreover, the problem
of rainfall or soil moisture retrieval from satellite microwave radiance,
can be considered as a non-linear inverse problem. This nonlinear
inversion may be cast in the presented context, provided that the
nonlinear kernel can be (locally) linearized with sufficient accuracy.
Application of regularization in the DA problems is in its infancy
\citep[e.g.; see,][for a recent study]{[FreNB12]} and is expected
to play significant role over the next decades, especially for highly
non-linear chaotic dynamical systems with frequent jumps.

\section*{Acknowledgment}

This work has been supported by an Interdisciplinary Doctoral Fellowship
(IDF) of the University of Minnesota Graduate School and the NASA-GPM
award NNX07AD33G. Partial support by a NASA Earth and Space Science
Fellowship (NESSF-NNX12AN45H) to the first author and the Ling chaired
professorship to the second author are also greatly acknowledged.
Thanks also go to Dr. Arthur Hou and Sara Zhang at NASA-Goddard Space
Flight Center for their support and insightful discussions.

\section*{Appendix}

\appendix
\numberwithin{equation}{section}

\section{Gradient Projection for Huber regularization }

Here, we present the gradient project (GP) method, using the Huber
regularization, only for the downscaling (DS) problem, which can be
easily generalized to the data fusion (DF) and data assimilation (DA)
cases. In case of the DS problem, the cost function and gradient of
the Huber regularization with respect to the elements of the downscaled
field are
\begin{equation}
\mathcal{J}(\mathbf{x})=\frac{1}{2}\left\Vert \mathbf{y}-\mathbf{Hx}\right\Vert _{\mathbf{R}^{-1}}^{2}+\lambda\,\left\Vert \mathbf{L}\mathbf{x}\right\Vert _{{\rm Hub}}\label{eq:A1}
\end{equation}
\begin{equation}
\nabla\mathcal{J}(\mathbf{x})=\mathbf{H}^{T}\mathbf{R}^{-1}\left(\mathbf{y}-\mathbf{H}\mathbf{x}\right)+\lambda\mathbf{L}^{T}\rho_{T}^{\prime}\left(\mathbf{Lx}\right),
\end{equation}
where
\begin{equation}
\rho_{T}^{\prime}\left(x\right)=\begin{cases}
2x & \left|x\right|\leq T\\
2T{\rm sign}(x), & \left|x\right|>T.
\end{cases}
\end{equation}
As is evident, the cost function in (\ref{eq:A1}), is a smooth and
convex function. Thus its minimum can be easily obtained using efficient
first order gradient descent methods in large dimensional problems.
However, rainfall is a positive process and in order to obtain a feasible
downscaled field $\hat{\mathbf{x}}$, the regularized DS problem needs
to be solved on the non-negative orthant $\{\mathbf{x}|\,\, x_{i}\geq0\,\,\forall\, i=1,\ldots,m\}$,
\begin{eqnarray}
\hat{\mathbf{x}} & = & {\rm argmin}\,\left\{ \mathcal{J}(\mathbf{x})\right\} \nonumber \\
 &  & \,\,{\rm s.t.\,\,\mathbf{x}\succeq0}.\label{eq:A4}
\end{eqnarray}
We have used one of the primitive gradient projection (GP) methods
to solve the above constrained DS problem \citep[see,][pp. 228]{[Ber99]}.
Accordingly, to obtain the solution of (\ref{eq:A4}), it amounts
to obtaining the fixed point of
\begin{equation}
\mathbf{x}_{*}=\left[\mathbf{x}_{*}-\alpha\nabla\mathcal{J}(\mathbf{x}_{*})\right]^{+},
\end{equation}
where $\alpha$ is a stepsize and
\begin{equation}
\left[x\right]^{+}=\begin{cases}
0 & \,\,{\rm if}\,\, x\leq0\\
x & {\rm otherwise,}
\end{cases}\label{eq:A6}
\end{equation}
denotes the Euclidean projection operator onto the non-negative orthant.
As is evident, the fixed point can be obtained iteratively as
\begin{equation}
\mathbf{x}_{k+1}=\left[\mathbf{x}_{k}-\alpha_{k}\nabla\mathcal{J}(\mathbf{x}_{k})\right]^{+}.
\end{equation}
Thus, if the descent at step $k$ is feasible (i.e., $\mathbf{x}_{k}-\alpha_{k}\nabla\mathcal{J}(\mathbf{x}_{k})\succeq0$)
the GP iteration becomes an ordinary unconstrained steepest descent
method, otherwise the result is mapped back onto the feasible set
by the projection operator in (\ref{eq:A6}). In effect, the GP method
finds iteratively the closest feasible point, in the Euclidean distance,
to the solution of the original unconstrained minimization.

In our study, the stepsize ($\alpha_{k}$) was selected using the\emph{
Armijo rule,} or the so called\emph{ backtracking line search}, that
is a convergent and very effective stepsize rule and depends on two
constants $0<\xi<0.5$ , $0<\varsigma<1$. In this method, the step
size is assumed $\alpha_{k}=\varsigma^{m_{k}}$, where $m_{k}$ is
the smallest non-negative integer for which
\begin{equation}
\mathcal{J}\left(\mathbf{x}_{k}-\alpha_{k}\nabla\mathcal{J}(\mathbf{x}_{k})\right)\leq\mathcal{J}(\mathbf{x}_{k})-\xi\alpha_{k}\nabla\mathcal{J}(\mathbf{x}_{k})^{T}\nabla\mathcal{J}(\mathbf{x}_{k}).
\end{equation}
In our DS examples the above backtracking parameters are set to $\xi=0.2$
and $\varsigma=0.5$ \citep[see,][pp.464 for more explanation]{[BoyV04]}.
In our coding, the iterations terminate if $\frac{\left\Vert \mathbf{x}_{k}-\mathbf{x}_{k-1}\right\Vert _{2}}{\left\Vert \mathbf{x}_{k-1}\right\Vert _{2}}\leq10^{-5}$
or the number of iterations exceeds 200.

\noindent {\small \bibliographystyle{agu04}
\addcontentsline{toc}{section}{\refname}\bibliography{MyRefs}
}
\end{document}